\documentclass[a4paper,10pt]{article}
\usepackage[utf8]{inputenc}
\usepackage[top=2.5cm,bottom=2.5cm,left=2.5cm,right=2.5cm]{geometry}
\usepackage{amsmath}
\usepackage{amssymb}
\usepackage{amsthm}
\usepackage{graphicx}
\usepackage{epstopdf}

\begin{document}

\title{Projection-free approximate balanced truncation \\ of large unstable systems}
\author{Thibault L. B. Flinois\footnote{Corresponding author: t.flinois11@imperial.ac.uk} \footnote{Department of Aeronautics, Imperial College London, South Kensington Campus, London SW7 2AZ, UK}, Aimee S. Morgans\footnotemark[\value{footnote}], Peter J. Schmid\footnote{Department of Mathematics, Imperial College London, South Kensington Campus, London SW7 2AZ, UK}}

\date{\today}
\maketitle

\begin{abstract}
In this article, we show that the projection-free, snapshot-based, balanced truncation method can be applied directly to unstable systems. We prove that even for unstable systems, the unmodified balanced proper orthogonal decomposition algorithm theoretically yields a converged transformation that balances the Gramians (including the unstable subspace). We then apply the method to a spatially developing unstable system and show that it results in reduced-order models of similar quality to the ones obtained with existing methods. Due to the unbounded growth of unstable modes, a practical restriction on the final impulse response simulation time appears, which can be adjusted depending on the desired order of the reduced-order model. Recommendations are given to further reduce the cost of the method if the system is large and to improve the performance of the method if it does not yield acceptable results in its unmodified form. Finally, the method is applied to the linearized flow around a cylinder at Re = 100 to show that it actually is able to accurately reproduce impulse responses for more realistic unstable large-scale systems in practice. The well-established approximate balanced truncation numerical framework can therefore be safely applied to unstable systems without any modifications. Additionally, balanced reduced-order models can readily be obtained even for large systems, where the computational cost of existing methods is prohibitive.\end{abstract}


\section{Introduction}
Many linear dynamical systems, such as the linearized Navier-Stokes equations, are composed of a large number of states ${\it O}(10^5 - 10^8)$, but their behavior is dominated by a much smaller number of modes ${\it O}(1 - 100)$. Obtaining a low-order model that only retains the dominant features of the system's behavior is of great value in order to understand and modify its dynamics. Specifically in a feedback control setting, most controller design methods are only tractable if a reduced-order model (ROM) is available.

In fluid mechanics, many successful ROMs of the flow field dynamics have been obtained by projecting the system equations onto a low-dimensional subspace, composed of a set of particularly relevant modes, such as global modes, proper orthogonal modes (POMs) or balanced modes. Global modes are ranked by their damping rate, so projecting the dynamics onto the least stable modes is a natural choice. {\AA}kervik {\it et al.}  \cite{Akervik2007} and Henningson {\it et al.} \cite{Henningson2008} successfully applied this strategy to damp global oscillations in a shallow cavity using an LQG controller. However, it is common for the dynamics to be dominated by the non-normal interaction between (potentially highly damped) global modes, in which case a large number of modes may be required to obtain an acceptable ROM. A more strategic selection of global modes can result in better performance in some cases \cite{Ehrenstein2010, Barbagallo2011} but finding a set of robust selection criteria is not straightforward \cite{Barbagallo2011}. Another weakness of this method is that identifying a large number of (highly damped) global modes may be prohibitively expensive.

Alternatively, proper orthogonal modes provide an optimally low rank approximation of the state over a chosen set of snapshots as they are ranked by their energy content. ROMs of many flow-fields have been developed by projecting the dynamics onto these modes. For instance, models of boundary layers \cite{Aubry1988,  Prabhu2001}, channel flows \cite{Podvin1998}, backward-facing steps \cite{Ravindran2000,  Ravindran2000a, Barbagallo2012}, bluff body flows and wakes \cite{Graham1999, Ma2002, Noack2003, Ravindran2006, Siegel2008, Tadmor2010}, forward facing steps \cite{Ravindran2002}, and cavities  \cite{Gloerfelt2008} have all been studied. POD-based ROMs are attractive both for the simplicity of the (snapshot-based) method used to obtain them \cite{Sirovich1987}, and the intuitive projection basis they provide (which allows retaining most of the energy of the snapshot ensemble). However, there is not a strong theoretical justification for using this orthogonal basis, as it depends on the specific set of snapshots used  to construct it. It is therefore unsurprising that a large number of modes are sometimes required to accurately represent the flow dynamics \cite{Barbagallo2009}. Several methods have been developed to improve the robustness and reliability of the models, such as regularly adapting the set of snapshots used to form the basis \cite{Ravindran2000a, Bergmann2008}, or adding a so called {\it shift-mode} \cite{Noack2003} which acts as a mean flow-correction mode and allows transients to be modeled more accurately.

In this article, we focus instead on ROMs based on balanced modes. These are ranked by their dynamical significance to the input-output relationship of the system, and are therefore better suited for feedback control purposes than POMs or global modes by design. ROMs based on balanced modes have attracted a significant amount of attention in recent years, and have been used in many flow configurations such as the response of the vertical force of an airfoil to a plunging motion \cite{Willcox2002}, a channel flow \cite{Rowley2005a}, a cavity flow \cite{Barbagallo2009}, a boundary layer flow \cite{Bagheri2009a, Bagheri2009b}, the flow over a flat plate at large incidence \cite{Ahuja2010}, over a backward facing step \cite{Dergham2011}, in a three dimensional boundary layer \cite{Semeraro2011}, and over a cylinder \cite{Tu2012}. Furthermore, several studies have compared the performance of ROMs based on global, proper orthogonal and balanced modes \cite{Willcox2002, Barbagallo2009, Bagheri2009c, Dergham2011} and have concluded that balanced ROMs typically require a much smaller number of modes for a given degree of accuracy.

As the classical (exact) balanced truncation method is intractable for large systems, Moore \cite{Moore1981}, Willcox and Peraire  \cite{Willcox2002}, and Rowley \cite{Rowley2005a}, developed an approximate snapshot-based method sometimes called balanced POD (or BPOD, see Section  \ref{baltruncstable}) to reduce the computational cost of the balancing procedure. When using this snapshot method, approximate balanced modes can be obtained from as few as two impulse response simulations (for single-input-single-output or SISO systems), and the snapshots do not need to be updated or complemented by a shift-mode. 

A further issue with the classical and snapshot methods is that they were designed for stable systems. While a number of studies \cite{Kenney1987, Meyer1990, Enns1985, Zhou1999, Dergham2011} have developed techniques to balance unstable systems (see \ref{baltruncunstable}), only a recent extension of the snapshot method \cite{Barbagallo2009, Ahuja2010} has allowed the computation of ROMs for {\it large} unstable systems. The method was applied to the flow over a cavity \cite{Barbagallo2009}, a flat plate at large incidence \cite{Ahuja2010}, and a cylinder \cite{Tu2012}. This extension requires the computation of the system's unstable global modes and the projection of the system onto its stable subspace. For large systems, such as three-dimensional flows, the cost of this procedure may still be excessively large.

The aim of this paper is to show that this expensive projection step is in fact unnecessary and that applying the unmodified snapshot-based balanced truncation method (designed for stable systems) directly to unstable systems yields a converged transformation, which balances the system and can result in ROMs of the same quality as those obtained when the projection method is used. Many large-scale unstable systems (e.g. three-dimensional flows) for which the projection step makes approximate balanced truncation intractable can therefore readily obtain balanced reduced-order models by using the projection-free method. 

In Sec.~\ref{baltrunc}, the theoretical framework is introduced and the existing balanced truncation methods are outlined. The fact that the projection-free snapshot method can be used for unstable systems is proven in Sec.~\ref{newbaltrunc}, and it is then applied to a representative one-dimensional model system in Sec.~\ref{applitogle}, where it is compared to existing methods. In Sec.~\ref{applitons}, the method is applied to a two-dimensional Navier-Stokes simulation of the flow over a cylinder to show that it also performs well in this more realistic and computationally demanding setup.


\section{Background \label{baltrunc}}
\subsection{Balanced truncation of stable systems \label{baltruncstable}}
In this section, we introduce the main concepts and methods that are relevant to balanced truncation, in particular for stable systems. More details about these methods can be found in standard control textbooks (e.g. \cite{Zhou1996} or \cite{Antoulas2005}). The standard continuous time, linear time-independent state-space system is:
\begin{eqnarray}
\label{eq:syscont}
\begin{cases}
\dot{x}=Ax+Bu,\\
y=Cx,
\end{cases}
\end{eqnarray}
where $x\in\mathbb{C}^{n_x}$ are the states of the system, $u\in\mathbb{C}^{n_u}$ are the inputs, $y\in\mathbb{C}^{n_y}$ are the outputs and $A\in\mathbb{C}^{n_x\times n_x}$, $B\in\mathbb{C}^{n_x\times n_u}$ and $C\in\mathbb{C}^{n_y\times n_x}$ are three time independent matrices. The dynamics of the system are then governed by the {\it transfer matrix}: $G(s)=C\left(sI-A\right)^{-1}B$, where $Y(s)=G(s)U(s)$, $s$ is the Laplace variable and $U(s)$ and $Y(s)$ are the Laplace transforms of the input and output signals, respectively.

The balanced truncation approach is based on an analysis of the {\it controllability} and {\it observability} of the system. The controllability of a state is related to the minimum input energy required to reach it from $x(0)=0$. The observability of a state $x_0$ is related to the energy of the output signal generated by the system starting from $x(0)=x_0$, without any input: $u(t)=0$. A state that has a large impact on the input-output behavior of the system - i.e. on the transfer matrix - is said to be {\it dynamically significant}. 

If a state is unobservable or nearly unobservable, then even if only a small amount of input energy is required to reach it (i.e. if it is highly controllable), it will not have a large impact on the output signal. Conversely, if a state is uncontrollable or nearly uncontrollable, then a large (or infinite) amount of input energy is required to reach it, so only a comparatively negligible part the output energy can be due to that state (as long as $x(0)=0$). Balanced truncation therefore aims to create a ROM by only retaining the states whose observability {\it and} controllability is high, as these are the ones which have the largest impact on the system dynamics.  In order to identify which states are the most controllable and observable, let us define the {\it controllability Gramian}:
\begin{eqnarray}
\label{eq:Wc}
W_c(t_\infty)=\int_0^{t_\infty}e^{At}BB^\dagger e^{A^\dagger t}dt,
\end{eqnarray}
and the {\it observability Gramian}:
\begin{eqnarray}
\label{eq:Wo}
W_o(t_\infty)&=&\int_0^{t_\infty}e^{A^\dagger t}C^\dagger Ce^{At}dt,
\end{eqnarray}
where $^\dagger $ is the complex conjugate transpose. These are defined for stable and unstable systems for ${0\le t_\infty<+\infty}$. As $t_\infty\to+\infty$ however, the Gramians converge to constant matrices for stable systems but become unbounded for unstable systems. When the limits exist, the following notation is used: ${\lim_{t_\infty\to+\infty}W_o(t_\infty)= W_o}$ and ${\lim_{t_\infty\to+\infty}W_c(t_\infty)= W_c}$.

A stable system is said to be {\it balanced} when ${W_c=W_o=\Sigma^2}$ where ${\Sigma=\text{diag}\left(\left[\begin{array}{ccc}\sigma_1&\ldots&\sigma_{n_x}\end{array}\right]\right)}$ and ${\Sigma\in\mathbb{R}^{n_x\times n_x}}$ is a diagonal matrix, where ${\sigma_1\ge\sigma_2\ge\ldots\ge\sigma_{n_x}}$ are the {\it Hankel singular values} (HSVs) of the system. The (non-zero) HSVs are unique and provide an indication of the corresponding state's dynamical significance. In other words, the states of a balanced system are ranked according to their joint controllability and observability. 

In general however, systems are not balanced and given a transfer matrix $G(s)$, the {\it realization} $(A,B,C)$ or internal coordinate system used to define the states $x$ is not unique. In order to balance a system it is therefore necessary to find the coordinate transformation ${T=S^{-1}\in\mathbb{C}^{n_x\times n_x}}$ that ensures: ${SW_cS^\dagger =T^\dagger W_oT=\Sigma^2}$. The transformed (balanced) system is then:
\begin{eqnarray}
\label{eq:balsys}
\begin{cases}
\dot{\hat{x}}=SAT\hat{x}+SBu,\\
y=CT\hat{x}.
\end{cases}
\end{eqnarray}
In order to obtain $T$ and $S$ for stable systems, the converged Gramians are evaluated by solving the following Lyapunov equations:
\begin{eqnarray}
\label{eq:lyaps}
\begin{cases}
A^\dagger W_o+W_oA+C^\dagger C=0,\\
AW_c+W_cA^\dagger +BB^\dagger =0.
\end{cases}
\end{eqnarray}
The balancing transformations are then found using Eq.~(\ref{eq:baltransTS}), which can be computed from the singular value decompositions (SVDs) in Eq.~(\ref{eq:svdWcWo}) and (\ref{eq:svdMH}):
\begin{subequations}
\begin{align}
&W_c=XX^\dagger \label{eq:svdWcWo},\quad W_o=ZZ^\dagger, \\
&Z^\dagger X=U\Sigma V^\dagger =\left[\begin{array}{cc}U_1&U_2\end{array}\right]\left[\begin{array}{cc}\Sigma_1&0\\0&0\end{array}\right]\left[\begin{array}{c}V_1^\dagger \\V_2^\dagger \end{array}\right],\label{eq:svdMH}\\
&T=XV_1\Sigma_1^{-1/2},\quad S=\Sigma_1^{-1/2}U_1^\dagger Z^\dagger \label{eq:baltransTS}.
\end{align}
\end{subequations}
Note that $X$ and $Z$ are not unique. The balanced system can then be decomposed as follows:
\begin{eqnarray}
\begin{cases}
\left[\begin{array}{c}\dot{\hat{x}}_1\\\dot{\hat{x}}_2\end{array}\right]=
\left[\begin{array}{cc} \hat{A}_{11}&\hat{A}_{12}\\\hat{A}_{21}&\hat{A}_{22}\end{array}\right]
\left[\begin{array}{c}\hat{x}_1\\\hat{x}_2\end{array}\right]+
\left[\begin{array}{c}\hat{B}_1\\\hat{B}_2\end{array}\right]u,\\\\
y=\left[\begin{array}{cc}\hat{C}_1&\hat{C}_2\end{array}\right]
\left[\begin{array}{c}\hat{x}_1\\\hat{x}_2\end{array}\right],
\end{cases}\nonumber
\end{eqnarray}
and as the transformed states are now ranked by dynamical significance, the less observable and controllable states $\hat{x}_2$ can be truncated to give the following ROM:
\begin{eqnarray}
\begin{cases}
\dot{\hat{x}}_1=\hat{A}_{11}\hat{x}_1+\hat{B}_1u,\\
y=\hat{C}_1\hat{x}_1,
\end{cases}\nonumber
\end{eqnarray}
which is also a stable balanced system. An attractive property of balanced truncation is that an {\it a priori} upper-bound on the ROM error exists: ${\lVert G-G_r \rVert_\infty<2\sum_{i=r+1}^{n_x}\sigma_i}$, where $r$ is the number of states that have been retained in the ROM, $\lVert \cdot \rVert_\infty$ is the $\cal{H}_\infty$ norm, and $G_r(s)=\hat{C}_1(sI-\hat{A}_{11})^{-1}\hat{B}_1$ is the reduced transfer matrix.

Unfortunately, if the system matrices are not explicitly available or if the system dimension is too large, this technique cannot be applied directly. Moore \cite{Moore1981}, Willcox and Peraire \cite{Willcox2002} and Rowley \cite{Rowley2005a} therefore introduced an approximate balanced truncation method sometimes referred to as balanced POD (BPOD), which is based on snapshots from the impulse response of the primal system (\ref{eq:syscont}) and the adjoint system (\ref{eq:sysadjcont}), also assumed to be available:
\begin{eqnarray}
\label{eq:sysadjcont}
\begin{cases}
\dot{z}=A^\dagger z+C^\dagger v,\\
w=B^\dagger z.
\end{cases}
\end{eqnarray}
In this approach, snapshots from the primal impulse responses ${x(t_{ck})=x_k=e^{\left(At_{ck}\right)}B}$ and the adjoint impulse response ${z(t_{ok})=z_k=e^{\left(A^\dagger t_{ok}\right)}C^\dagger }$ are stacked into matrices in order to find an approximation for the Gramians in the form Eq.~(\ref{eq:svdWcWo}), by defining $X$ and $Z$ in the following way:
\begin{subequations}\label{eq:xzapprox}\begin{align}
&X=\left[\begin{array}{ccc}x_1\sqrt{\delta_{c1}}&\ldots&x_{N_c}\sqrt{\delta_{cN_c}} \end{array}\right]\Rightarrow W_c\approx XX^\dagger,\\
&Z=\left[\begin{array}{ccc}z_1\sqrt{\delta_{o1}}&\ldots&z_{N_o}\sqrt{\delta_{oN_o}} \end{array}\right]\Rightarrow W_o\approx ZZ^\dagger,
\end{align}\end{subequations}
where there are $N_c$ primal snapshots and $N_o$ adjoint snapshots taken at discrete times $t_{ck}$ and $t_{ok}$ respectively. $\delta_{ck}\in\mathbb{R}$ and $\delta_{ok}\in\mathbb{R}$ are the associated quadrature coefficients corresponding to a chosen numerical integration scheme. As a result, $XX^\dagger$ and $ZZ^\dagger$ are discrete versions of the continuous definition of the Gramians in Eq.~(\ref{eq:Wc}) and Eq.~(\ref{eq:Wo}), respectively. Note that $x_k\in\mathbb{C}^{n_x\times n_u}$ and $z_k\in\mathbb{C}^{n_x\times n_y}$ and that the snapshots are not necessarily equally spaced in time. Forming $X$ and $Z$ in this manner therefore allows $T$ and $S$ to be computed using Eq.~(\ref{eq:baltransTS}), but at a reduced cost since it avoids finding the solution of the Lyapunov equations (\ref{eq:lyaps}) and also only requires finding the SVD of a $(N_cn_u\times N_on_y)$ matrix instead of the three $(n_x\times n_x)$ matrices in Eq.~(\ref{eq:svdWcWo}) and (\ref{eq:svdMH}). This can represent significant savings if $N_cn_u\ll n_x$ and $N_on_y\ll n_x$, which is usually the case in computational fluid dynamics. Additionally, ``squaring up'' the matrices $X$ and $Z$ to form the Gramians is detrimental to the accuracy of the results \cite{Moore1981}.

\subsection{Balanced truncation of unstable systems \label{baltruncunstable}}
In this section, existing methods related to the balanced truncation of unstable systems are discussed. For stable systems it is straightforward to show that the balancing transformation $T$ converges and that it diagonalizes and equalizes the Gramians as $t_\infty\to+\infty$, since the primal and adjoint states go to zero. If the system is unstable however the state becomes unbounded for large $t_\infty$. $W_c(t_\infty)$, $W_o(t_\infty)$ as defined in Eq.~(\ref{eq:Wc}) and (\ref{eq:Wo}) respectively are therefore also unbounded and $W_c$ and $W_o$ are not defined.

Nevertheless, Chiu \cite{Chiu1996}, Kenney and Hewer \cite{Kenney1987}, Therapos  \cite{Therapos1989}, and Al-Saggaf \cite{Al-Saggaf1992} showed that despite Eq.~(\ref{eq:Wc}) and (\ref{eq:Wo}) being unbounded, Eq.~(\ref{eq:lyaps}) still have solutions if and only if $\lambda_i+\lambda_j\neq0$, where $\lambda_i$ and $\lambda_j$ are any two eigenvalues of $A$. When this is the case, a balancing transformation can be obtained if $W_cW_o$ is similar to a real diagonal matrix. However, this is problematic when Eq.~(\ref{eq:lyaps}) does not have a solution and when the balancing transformation does not exist. Furthermore, the resulting reduced-order models are not always of satisfactory quality \cite{Zhou1999}.

An alternative method developed by Meyer \cite{Meyer1990} is based on the co-prime factorization of the transfer matrix. Nett \cite{Nett1984} showed that ${G(s)=N(s)M(s)^{-1}}$, where ${N(s)=C(sI-\bar{A})^{-1}B}$ and $M(s)=I+K(sI-\bar{A})^{-1}B$ is a right co-prime factorization of $G(s)$ if ${\bar{A}=A+BK}$ is stable. Meyer and Franklin \cite{Meyer1987} showed that if ${K=-B^\dagger P}$, where $P$ is the solution to the algebraic Riccati equation: ${PA+A^\dagger P-PBB^\dagger P+C^\dagger C=0}$ then the co-prime factorization is normalized. A realization of the stable system ${\left[\!\!\begin{array}{cc}N^\dagger (s)&M^\dagger (s)\end{array}\!\!\right]^\dagger }$ is then given by ${\left(\bar{A},B,\left(\!\!\begin{array}{cc}C^\dagger &K^\dagger \end{array}\!\!\right)^\dagger ,\left(\!\!\begin{array}{cc}0&I\end{array}\!\!\right)^\dagger \right)}$, which can be balanced and truncated as in Sec.~\ref{baltruncstable}. The stabilizing feedback $K$ and output augmentation can then be undone to retrieve the ROM. An issue with this method is that it does not simplify to the standard method when the plant is stable.

Zhou \cite{Zhou1999} developed a related method by using the frequency-domain definition of the Gramians:
\begin{eqnarray}
W_{cf}=\frac{1}{2\pi}\int_{-\infty}^{+\infty}\left(i\omega I-A\right)^{-1}BB^\dagger \left(-i\omega I-A^\dagger \right)^{-1}d\omega\nonumber,\\
W_{of}=\frac{1}{2\pi}\int_{-\infty}^{+\infty}\left(-i\omega I-A^\dagger \right)^{-1}C^\dagger C\left(i\omega I-A\right)^{-1}d\omega\nonumber,
\end{eqnarray}
where $W_{cf}$ and $W_{of}$ are now well-defined as long as $A$ has no eigenvalues on the imaginary axis. In the stable case, Parseval's  theorem can be applied to show that ${W_{cf}=W_c}$ and ${W_{of}=W_o}$. It is then possible to evaluate $W_{cf}$ and $W_{of}$ if a transformation that decouples the stable and antistable dynamics of the system is available. However, they are usually instead evaluated by finding the stabilizing solutions to ${(A+BK)W_{cf}+W_{cf}(A+BK)^\dagger +BB^\dagger =0}$ and ${W_{of}(A+LC)+(A+LC)^\dagger W_{of}+C^\dagger C=0}$, where ${K=-B^\dagger P}$ and $L=-YC$. Here $P$ is the stabilizing solution to the algebraic Riccati equation ${PA+A^\dagger P-PBB^\dagger P=0}$ and $Y$ is the stabilizing solution to the algebraic Riccati equation $AY+YA^\dagger -YC^\dagger CY=0$. The balancing and truncation procedure from Sec.~\ref{baltruncstable} is applied once these ``generalized'' Gramians are known.

A snapshot-based extension to this method was developed by Dergham {\it et al.}  \cite{Dergham2011}, who applied it to a rounded backward facing step and cavity flow. Here, the snapshots are defined in the frequency-domain rather than the time-domain: the primal system snapshots are of the form ${X\left(\omega\right)=\left(i\omega I-A\right)^{-1}B}$ and of an analogous form for the adjoint snapshots. Each snapshot therefore has to be evaluated by inverting a large matrix explicitly, but all the snapshots can potentially be evaluated simultaneously in parallel.

In the methods described above, the Hankel singular values corresponding to unstable modes are not necessarily larger than the ones corresponding to stable modes, so there is no guarantee that unstable modes will not be truncated. This is often not a desired property for control design purposes. If on the other hand the transformation that uncouples the stable and antistable dynamics is available, such that ${G(s)=G_a(s)+G_s(s)}$, then an alternative approach, for instance suggested by Enns \cite{Enns1985}, is to only balance and truncate the stable part of the system $G_s(s)$ and simply add the unbalanced antistable dynamics back in such that the ROM transfer matrix is: ${G_r=G_a(s)+\hat{G}_s(s)}$. where $\hat{\cdot}$ refers to the balancing and truncation procedure, thus conserving all the unstable modes.

Unfortunately, if only a time-stepping code is available and if the system dimension is large, none of the methods described above can be used. In order to tackle this issue, an extension to the snapshot method introduced in Sec.~\ref{baltruncstable} was developed by Barbagallo {\it et al.}  \cite{Barbagallo2009} and Ahuja {\it et al.}   \cite{Ahuja2010}. This extension is closely related to the approach proposed by Enns  \cite{Enns1985} as it is based on the separation of the stable and antistable dynamics of the system. Although it may be expensive to compute the full stable-antistable uncoupling transformation, it may still be possible to use an Arnoldi procedure to identify the right and left antistable eigenspaces $P_a$ and $Q_a$ respectively (scaled such that ${Q_a^\dagger P_a=I}$). The primal and adjoint systems can then be projected onto their respective stable subspaces, by defining the projection matrix ${{\cal P}=I-P_aQ_a^\dagger }$:
\begin{eqnarray}
\begin{cases}
\dot{x}_s={\cal P}A{\cal P}x_s+{\cal P}Bu,\\
y_s=C{\cal P}x_s,\nonumber
\end{cases}
\end{eqnarray}
and
\begin{eqnarray}
\begin{cases}
\dot{z}_s={\cal P}^\dagger A^\dagger {\cal P}^\dagger z_s+{\cal P}^\dagger C^\dagger v,\\
w_s=B^\dagger {\cal P}^\dagger z_s\nonumber.
\end{cases}
\end{eqnarray}
This projected system is then balanced and truncated using snapshots, as described in Sec.~\ref{baltruncstable}. The antistable dynamics are finally added back in, and the final reduced-order model is:
\begin{eqnarray}
\begin{cases}
\dot{x_r}=A_rx_r+B_ru,\\
y=C_rx_r,\end{cases}\nonumber
\end{eqnarray}
where
\begin{eqnarray}
A_r=\left[\begin{array}{cc}Q_a^\dagger AP_a&0\\0&\hat{A}\end{array}\right],
B_r=\left[\begin{array}{c}Q_a^\dagger B\\\hat{B}\end{array}\right],
C_r=\left[\begin{array}{cc}CP_a&\hat{C}\end{array}\right]\nonumber,
\end{eqnarray}
where, as above $\hat{\cdot}$ refers to the projected system's balanced and truncated matrices. Like in the method proposed by Enns \cite{Enns1985}, this procedure does not balance the unstable subspace (unlike the ones developed by Kenney and Hewer  \cite{Kenney1987}, Meyer \cite{Meyer1990}, and Zhou \cite{Zhou1999}), but guarantees that unstable modes are not truncated. For fluid systems, the dimension of the unstable subspace is often ${\it O}(1-10)$ so this method often only requires computing a few eigenvalues and eigenmodes. However an Arnoldi package is not always available and for large systems (e.g. three-dimensional flows), finding even a few eigenmodes may still not be tractable. A further issue with this method is that if the system is poorly conditioned, it may be difficult to evaluate the projection matrix itself.

In this article we propose an alternative method that is projection-free and does not require evaluating any global modes. As briefly mentioned in \cite{Antoulas2005}, one can choose to simply use the standard method for stable systems, but based on {\it finite-time} Gramians, which are expected to approximate the impulse response over the chosen time interval well-enough. This article aims to show that if this finite time interval is long enough, the balancing transformations actually converge and hence balance the Gramians for any further time. These transformations can therefore be considered to be true balancing transformations, often obtained at a fraction of the cost of the methods described above.  As a result, we show that the projection-free, snapshot-based balanced truncation method is directly applicable to unstable systems.


\section{Projection-free balanced truncation of unstable systems \label{newbaltrunc}}
\subsection{Theoretical justification}
The goal of this section is to show that the projection-free snapshot-based balanced truncation method (BPOD) introduced in Sec.~\ref{baltruncstable} can be used even in the unstable case, despite the fact that the Gramians are unbounded. In order to complete the proof we set out to prove the following statements:
\begin{enumerate}
 \item The transformations $T$ and $S$ as defined in Eq.~(\ref{eq:baltransTS}) converge to constant matrices, even for unstable systems.
 \item The controllability and observability Gramians are balanced by the converged transformations for any sufficiently large $t_\infty$, despite not converging to constant matrices.
\end{enumerate}
These two statements are proven in {\it Proposition 2} and {\it Proposition 3} respectively and ensure that the converged $T$ and $S$ matrices can be considered to be balancing transformations for the unstable system. These proofs are based on {\it Proposition 1}, which states that as ${t_\infty\to+\infty}$, the unstable singular vectors and values of the matrices $X$ and $Z$ can be identified explicitly. In particular, all the left singular vectors tend to constant vectors. The proofs outlined here correspond to the case where all eigenvalues have distinct real parts. Appendix~\ref{specialcases} deals with configurations where some of the modes have the same growth rate.

Let $(A,B,C)$ be a minimal realization, where ${A=P\Lambda Q^\dagger }$, and ${P^\dagger Q=I}$, where ${P=\left[\begin{array}{ccc}p_1&\ldots&p_{n_x}\end{array}\right]}$, ${\Lambda=\text{diag}\left(\left[\begin{array}{ccc}\lambda_1&\ldots&\lambda_{n_x}\end{array}\right]\right)}$, ${Q=\left[\begin{array}{ccc}q_1&\ldots&q_{n_x}\end{array}\right]}$, and ${\text{Re}(\lambda_1)<\ldots<\text{Re}(\lambda_{n_x})}$.  The primal and adjoint impulse responses are defined as in Sec.~\ref{baltruncstable} and can respectively be written:
\begin{eqnarray}
x(t_{ck})&=&x_k=Pe^{\left(\Lambda t_{ck}\right)}Q^\dagger B\nonumber,\\
z(t_{ok})&=&z_k=Qe^{\left(\Lambda^\dagger t_{ok}\right)}P^\dagger C^\dagger. \nonumber
\end{eqnarray}
The approximation of $X$ and $Z$ from the impulse response snapshots defined in Eq.~(\ref{eq:xzapprox}) can be written:
\begin{eqnarray}
 X= P\beta^\dagger =U_c\Sigma_cV_c^\dagger ,\qquad Z= Q\xi^\dagger =U_o\Sigma_oV_o^\dagger, \nonumber
\end{eqnarray}
where ${\beta=\left[\begin{array}{ccc}\beta_{1}&\ldots&\beta_{n_x}\end{array}\right]}$ and ${\xi=\left[\begin{array}{ccc}\xi_{1}&\ldots&\xi_{n_x}\end{array}\right]}$ and:
\begin{eqnarray}
\begin{array}{l}
\beta_{i}=\left[\begin{array}{c}\sqrt{\delta_{c1}}B^\dagger q_ie^{\left(\lambda_i^*t_{c1}\right)}\\\vdots\\\sqrt{\delta_{cN_c}}B^\dagger q_ie^{\left(\lambda_i^*t_{cN_c}\right)}\end{array}\right],\\\\
\xi_{i}=\left[\begin{array}{c}\sqrt{\delta_{o1}}Cp_ie^{\left(\lambda_it_{o1}\right)}\\\vdots\\\sqrt{\delta_{oN_o}}Cp_ie^{\left(\lambda_it_{oN_o}\right)}\end{array}\right],
\end{array}\nonumber
\end{eqnarray}
where the superscript $^*$ refers to the complex conjugate.

\bigskip\textbf{\textit{Proposition 1:} For large $t_\infty$, the singular values and vectors of $X$ ($Z$) corresponding to unstable singular values in $\Sigma_c$ ($\Sigma_o$) converge to:
\begin{subequations}
\label{eq:prop1}
\begin{align}
&\begin{cases}
u_{ci}=T_{ci-1}p_i\lVert T_{ci-1}p_i \rVert^{-1},\\
\sigma_{ci}=\lVert T_{ci-1}p_i \rVert\lVert \beta_{i} \rVert,\\
v_{ci}=\beta_{i}\lVert \beta_{i} \rVert^{-1},
\end{cases}\label{eq:prop1a}\\
&\begin{cases}
u_{oi}=T_{oi-1}q_i\lVert T_{oi-1}q_i \rVert^{-1},\\
\sigma_{oi}=\lVert T_{oi-1}q_i \rVert\lVert \xi_{i} \rVert,\\
v_{oi}=\xi_{i}\lVert \xi_{i} \rVert^{-1},
\end{cases}\label{eq:prop1b}
\end{align}
\end{subequations}
respectively, where $\lVert\cdot\rVert$ is the Euclidean norm of a vector and:
\begin{subequations}
\begin{align}
&T_{ci}=I-\left[\begin{array}{ccc}u_{c1}&\ldots&u_{ci}\end{array}\right]\left[\begin{array}{c}u_{c1}^\dagger \\\vdots\\u_{ci}^\dagger \end{array}\right]\label{eq:Tci},\\
&T_{oi}=I-\left[\begin{array}{ccc}u_{o1}&\ldots&u_{oi}\end{array}\right]\left[\begin{array}{c}u_{o1}^\dagger \\\vdots\\u_{oi}^\dagger \end{array}\right]\label{eq:Toi}.
\end{align}
\end{subequations}
}

\begin{proof}
We will use a proof by induction and choose the induction hypothesis $({\cal I}_n)$ to be the fact that the proof holds for all $i\le n$. 

For the base case $({\cal I}_1)$, if $\text{Re}(\lambda_1)>0$ then the first mode $p_1$ is unstable and:
\begin{eqnarray}
\lim_{k\to+\infty}x_k&=&p_1e^{\lambda_1t_{ck}}q_1^\dagger B,\nonumber\\
\lim_{t_\infty\to+\infty}X&=&p_1\beta^\dagger _{1}=u_{c1}\sigma_{c1}v_{c1}^\dagger, \nonumber
\end{eqnarray}
since this is just rank 1. $({\cal I}_{1})$ therefore holds, up to a unit norm factor $e^{i\theta_{c1}}$:
\begin{eqnarray}
\begin{cases}
u_{c1}= p_1\lVert p_1 \rVert^{-1}e^{i\theta_{c1}},\\
\sigma_{c1}=\lVert p_1 \rVert\lVert \beta_{1} \rVert,\\
v_{c1}= \beta_{1}\lVert \beta_{1} \rVert^{-1}e^{-i\theta_{c1}}\nonumber,
\end{cases}
\end{eqnarray}
where $\theta_{c1}$ is a real scalar. The first singular vectors and value are thus uniquely defined, up to $e^{i\theta_{c1}}$, which we can always choose to be equal to $1$, so we will make this assumption without loss of generality for the remainder of this paper to simplify the notation. As a result, the direction of the first left singular vector $u_{c1}$ converges to that of the first eigenvector of the system $p_1$ (which is constant).

In the inductive step, we prove that if $({\cal I}_n)$ holds for some rank $n$, then $({\cal I}_{n+1})$ also holds. The transformation matrices $T_{ci}$ defined in Eq.~(\ref{eq:Tci}) project out all the left singular vectors $u_{cj}$ for all $j\le i$, i.e. ${T_{ci}u_{cj}=0}$, ${T_{cj}u_{ci}=u_{ci}}$:
\begin{eqnarray}
T_{cn}X&=&\left[\begin{array}{ccc}u_{cn+1}&\ldots&u_{cn_x}\end{array}\right]\!\!
\left[\!\!\begin{array}{ccc}\sigma_{cn+1}&&\\&\ddots&\\&&\sigma_{cn_x}\end{array}\!\!\right]\!\!
\left[\!\!\begin{array}{c}v^\dagger _{cn+1}\\\vdots\\v^\dagger _{cn_x}\end{array}\!\!\right],\nonumber\\
&=&\left[\begin{array}{ccc}T_{cn}p_{1}&\ldots&T_{cn}p_{n}\end{array}\right]
\left[\begin{array}{c}\beta^\dagger _{1}\\\vdots\\\beta^\dagger _{n}\end{array}\right]\nonumber\\
&&+\left[\begin{array}{ccc}T_{cn}p_{n+1}&\ldots&T_{cn}p_{n_x}\end{array}\right]
\left[\begin{array}{c}\beta^\dagger _{n+1}\\\vdots\\\beta^\dagger _{n_x}\end{array}\right].\nonumber
\end{eqnarray}
Thus, if $p_{n+1}$ is an unstable mode and recalling that we assumed that ${\text{Re}(\lambda_1)<\ldots<\text{Re}(\lambda_{n_x})}$, we have:
\begin{eqnarray}
\lim_{t_\infty\to+\infty} T_{cn}X&=&\left[\begin{array}{ccc}T_{cn}p_{1}&\ldots&T_{cn}p_{cn}\end{array}\right]
\left[\begin{array}{c}\beta^\dagger _{1}\\\vdots\\\beta^\dagger _{n}\end{array}\right]\nonumber\\
&&+T_{cn}p_{n+1}\beta^\dagger _{n+1}\nonumber.
\end{eqnarray}
Now for $0<i\le n$, $T_{cn}$ can be written:
\begin{eqnarray}
T_{cn}&=&\left(I-\left[\begin{array}{ccc}u_{ci}&\ldots&u_{cn}\end{array}\right]\left[\begin{array}{c}u_{ci}^\dagger \\\vdots\\u_{cn}^\dagger \end{array}\right]\right)T_{ci-1}\nonumber,
\end{eqnarray}
and since we are assuming $({\cal I}_{n})$ holds for rank $n$:
\begin{eqnarray}
&&T_{cn}p_i=\!\left(I-\left[\!\!\begin{array}{ccc}u_{ci}&\ldots&u_{cn}\end{array}\!\!\right]\left[\!\!\begin{array}{c}u_{ci}^\dagger \\\vdots\\u_{cn}^\dagger \end{array}\!\!\right]\right)u_{ci}\lVert T_{ci-1}p_i\rVert=0\nonumber,\\
&&\Rightarrow \lim_{t_\infty\to+\infty} T_{cn}X=T_{cn}p_{n+1}\beta^\dagger _{n+1}=u_{cn+1}\sigma_{cn+1}v_{cn+1}^\dagger \nonumber,
\end{eqnarray}
and hence $({\cal I}_{n+1})$ holds too. This completes the inductive step and along with the base step concludes the proof by induction. The singular values and vectors corresponding to unstable modes are therefore given by:
\begin{eqnarray}
\begin{cases}
u_{ci}= T_{ci-1}p_i\lVert T_{ci-1}p_i \rVert^{-1},\\
\sigma_{ci}=\lVert T_{ci-1}p_i \rVert\lVert \beta_{i} \rVert,\\
v_{ci}= \beta_{i}\lVert \beta_{i} \rVert^{-1}\nonumber.
\end{cases}
\end{eqnarray}
The singular vector $u_{ci}$ is therefore pointing in the direction of the component of $p_i$ that is orthogonal to the subspace defined by ${\left[\begin{array}{ccc}p_1&\ldots&p_{i-1}\end{array}\right]}$. The procedure that identifies the left unstable controllability singular vectors is therefore essentially a Gram-Schmidt process. An analogous derivation starting from the adjoint impulse response leads to:
\begin{eqnarray}
\begin{cases}
u_{oi}= T_{oi-1}q_i\lVert T_{oi-1}q_i \rVert^{-1},\\
\sigma_{oi}=\lVert T_{oi-1}q_i \rVert\lVert \xi_{i} \rVert,\\
v_{oi}= \xi_{i}\lVert \xi_{i} \rVert^{-1}\nonumber,
\end{cases}
\end{eqnarray}
completing the proof of {\it Proposition 1}.
\end{proof}

It is clear that the left singular vectors and singular values of $X$ and $Z$, when restricted to the stable subspace also converge to constants as $t_\infty\to+\infty$, (this is the basis for the snapshot method in \cite{Barbagallo2009} and  \cite{Ahuja2010}). Therefore all stable {\it and} unstable left singular vectors of $X$ and $Z$ tend to constants, i.e. $U_c$ and $U_o$ converge to constant matrices.

\bigskip\textbf{\textit{Proposition 2:} The balancing transformations $T$ and $S$ converge to constant matrices for large $t_\infty$.}
\begin{proof} Let us define the Hankel matrix $M_H$, noting that the SVD of the Gramians can be approximated with snapshots for any finite $t_\infty$: ${W_c(t_\infty)\approx XX^\dagger }$, ${W_o(t_\infty)\approx ZZ^\dagger }$:
\begin{eqnarray}
M_H=Z^\dagger X=U\Sigma V^\dagger =V_o\Sigma_oU_o^\dagger U_c\Sigma_cV_c^\dagger .
\end{eqnarray}
Using a similar reasoning to the one in the proof of {\it Proposition 1}, the $(i+1)$th set of Hankel singular values and vectors corresponding to each unstable mode of the system can be identified by projecting $M_H$ onto the subspace that is orthogonal to the left and right singular vectors corresponding to all unstable modes $j$ such that $j\le i$. By using a similar proof by induction to that in {\it Proposition 1}  (see Appendix~\ref{hankinduction} for the full derivation) and by defining the following two transformations:
\begin{subequations}
\label{eq:TliTri}
\begin{align}
&T_{li}=I-\left[\begin{array}{ccc}u_{1}&\ldots&u_{i}\end{array}\right]\left[\begin{array}{c}u_{1}^\dagger \\\vdots\\u_{i}^\dagger \end{array}\right],\label{eq:Tli}\\
&T_{ri}=I-\left[\begin{array}{ccc}v_{1}&\ldots&v_{i}\end{array}\right]\left[\begin{array}{c}v_{1}^\dagger \\\vdots\\v_{i}^\dagger \end{array}\right],\label{eq:Tri}
\end{align}
\end{subequations}
it can be shown that as $t_\infty\to+\infty$:
\begin{eqnarray}
\begin{cases}
u_i=v_{oi},\\
\sigma_i=\sigma_{oi}u_{oi}^\dagger u_{ci}\sigma_{ci},\\
v_i= v_{ci}.\label{eq:hanksingvals}
\end{cases}
\end{eqnarray}
Because the singular values corresponding to unstable modes tend to infinity for large $t_\infty$, we can separate the stable and antistable parts of $X$ and $M_H$ (denoted with the subscripts $_s$ and $_a$ respectively) as follows:
\begin{eqnarray}
\lim_{t_\infty\to+\infty}X&=&\left[\begin{array}{cc}U_{ca}&U_{cs}\end{array}\right]\left[\begin{array}{cc}\Sigma_{ca}&0\\0&\Sigma_{cs}\end{array}\right]\left[\begin{array}{c}V_{ca}^\dagger \\V_{cs}^\dagger \end{array}\right],\nonumber\\
\lim_{t_\infty\to+\infty}M_H &=&\left[\begin{array}{cc}U_{a}&U_{s}\end{array}\right]\left[\begin{array}{cc}\Sigma_{a}&0\\0&\Sigma_{s}\end{array}\right]\left[\begin{array}{c}V_{a}^\dagger \\V_{s}^\dagger \end{array}\right].\nonumber
\end{eqnarray}
Now, Eq.~(\ref{eq:hanksingvals}) implies that $V_{ca}^\dagger V_a=I$, $V_{ca}^\dagger V_s=0$, and $V_{cs}^\dagger V_a=0$. As a result, the transformation matrix $T$, as defined in Eq.~(\ref{eq:baltransTS}) becomes:
\begin{eqnarray}
\lim_{t_\infty\to+\infty}T&=&\left[\begin{array}{cc}U_{ca}\Sigma_{ca}\Sigma_a^{-1/2}&U_{cs}\Sigma_{cs}V_{cs}^\dagger V_s\Sigma_s^{-1/2}\end{array}\right],\nonumber\\
&=&\left[\begin{array}{cc}T_{a}&T_{s}\end{array}\right].\label{eq:Tconverged}
\end{eqnarray}
The $i$th column of the converged $T_a$ is therefore ${T_{ai}=u_{ci}\sqrt{\sigma^2_{ci}/\sigma_{i}}}$. The ratio of the singular values can be shown to tend to a constant (this fact is proven in Appendix~\ref{cstsigratios}) so the matrix $T$ converges to a constant matrix for large $t_\infty$ (since $T_s$ must also converge). An analogous argument can be made to show that $S$ converges to a constant matrix too, where ${\lim_{t_\infty\to+\infty}S=\left[\begin{array}{cc}S_a^\dagger &S_s^\dagger \end{array}\right]^\dagger }$, where ${S_a=\Sigma_a^{-1/2}\Sigma_{oa}U_{oa}^\dagger }$ and ${S_s=\Sigma_s^{-1/2}U_s^\dagger V_{os}\Sigma_{os}U_{os}^\dagger }$.  This completes the proof of {\it Proposition 2}.
\end{proof}

\bigskip\textbf{\textit{Proposition 3:} The converged balancing transformation $T$ ($S$) balances the Gramian $W_o$ ($W_c$) for sufficiently large $t_\infty$.}
\begin{proof}
If a (converged) transformation $T(t_1)$ is found, corresponding to a given set to snapshots with the final snapshot taken at ${t_\infty\!=\!t_1}$, we would like to check that it diagonalizes and equalizes the Gramians $W_c(t_2)$ and $W_o(t_2)$, corresponding to a different (but also sufficiently large) set of snapshots such that ${t_\infty\!=\!t_2}$. Using the notation ${M_{H(21)}\!=\!Z(t_2)^\dagger X(t_1)\!=\!U_{(21)}\Sigma_{(21)}V_{(21)}^\dagger }$, the transformed observability Gramian becomes:
\begin{eqnarray}
&&T^\dagger W_oT= T(t_1)^\dagger W_o(t_2)T(t_1),\nonumber\\
&&\approx\left(\Sigma_{(11)}^{-1/2}V_{(11)}^\dagger X^\dagger (t_1)\right)
Z(t_2)Z(t_2)^\dagger 
\left(X(t_1)V_{(11)}\Sigma_{(11)}^{-1/2}\right),\nonumber\\
&&\approx\Sigma_{(11)}^{-1/2}\left(V_{(11)}^\dagger V_{(21)}\right)\Sigma_{(21)}^2 \left(V_{(21)}^\dagger V_{(11)}\right)\Sigma_{(11)}^{-1/2},\label{eq:diagWo}
\end{eqnarray}
and $V_{(21)}$ can be obtained from the matrix ${M_{H(21)}=V_o(t_2)\left[\Sigma_o(t_2)U_o^\dagger (t_2)U_c(t_1)\Sigma_c(t_1)\right]V_c(t_1)^\dagger }$. 
By following the same procedure as in the proof of {\it Proposition 2}, we obtain:
\begin{eqnarray}
U_{a(21)}=V_{oa}(t_2), \qquad V_{a(21)}=V_{ca}(t_1)=V_{a(11)},\label{eq:convVanti}
\end{eqnarray}
where, as above, the $_a$ subscript refers to the antistable part of the matrix for large $t_\infty$ (i.e. large $t_1$ and $t_2$). For the stable part, given Eq.~(\ref{eq:convVanti}) and the fact that the stable modes decay for large $t_1$ and $t_2$, any additional snapshots cannot modify these subspaces so ${V_{s(21)}=V_{s(11)}}$. Finally:
\begin{eqnarray}
\lim_{t_\infty\to+\infty}V_{(11)}^\dagger V_{(21)}&=&I,\nonumber\\
\Rightarrow\lim_{t_\infty\to+\infty}T(t_1)^\dagger W_o(t_2)T(t_1)&\approx&\Sigma_{(11)}^{-1}\Sigma_{(21)}^2,\label{eq:prop3}
\end{eqnarray}
which is clearly diagonal. An analogous proof can be applied to the controllability Gramian and leads to ${S(t_1)W_c(t_2)S^\dagger (t_1)\approx\Sigma_{(11)}^{-1}\Sigma_{(21)}^2\approx T(t_1)^\dagger W_o(t_2)T(t_1)}
$.  This completes the proof of {\it Proposition 3}.
\end{proof}
In Appendix~\ref{specialcases}, it is shown that the same conclusions hold for unstable and marginally stable repeated eigenvalues, as well as {\it marginally} stable complex conjugate pairs. For {\it unstable} complex conjugate pairs, only the two-dimensional subspace spanned by the two corresponding balanced modes converges. This is all that we require, as it is usually not desirable to truncate only one of the two modes corresponding to an unstable pair of complex conjugate eigenvalues. In this case, the two columns (rows) of the transformation $T$ ($S$) that define this converged two-dimensional subspace oscillate as $t_\infty$ is increased. The transformed Gramians are therefore equal and diagonal except for one $(2\times2)$ block along the diagonal for each unstable complex conjugate pair. Let us emphasize here that despite the transformation not becoming constant and the Gramians not being strictly diagonalized and equal, any $T$ (and $S$) corresponding to a sufficiently large $t_\infty$ can be used as the balancing transformation, as long as the 2D subspace corresponding to the unstable modes has converged. In this case the unstable complex conjugate mode pair is balanced as a whole.

\subsection{Practical considerations \label{practical}}
As illustrated in sections \ref{applitogle} and \ref{applitons}, the projection-free snapshot-based balanced truncation method can often be applied directly to large unstable systems, just as in the stable case. Nevertheless, some modifications to the method that can lead to significant improvements in the quality of the ROMs and/or the computational cost of the method are outlined in this section. These may be required for particularly challenging systems.

\subsubsection{Final simulation time  and sampling intervals\label{choosingsnaps}}
For both stable and unstable systems, the sampling intervals must be small enough to capture the highest frequencies of interest in the flow field and the final simulation time $t_\infty$ must be large enough for all stable modes to decay. For unstable systems $t_\infty$ must also be large enough for the impulse response to be dominated by the unstable modes at the end of the simulation, thus allowing $T_a$ and $S_a$ to converge.

Although there is no theoretical upper-bound on $t_\infty$, in practice as $t_\infty\to+\infty$ any initial transients eventually become so negligible compared to the long term response that the information related to modes that are more stable than the dominating unstable mode(s) may be lost. This can result in an inaccurate identification of the corresponding balanced modes. Both the upper and lower bounds on $t_\infty$ are clearly problem-dependent and the trade-off between these two limits is investigated in further detail in Sec.~\ref{applitogle}.

\subsubsection{Improving the accuracy of the method}
If several unstable modes have nearly identical growth rates or if it is necessary to identify slowly decaying modes, the lower bound on $t_\infty$ may be higher than its upper bound (see \ref{choosingsnaps}). It may therefore not be possible to obtain sufficiently accurate information about the less unstable parts of the system from the impulse response. One way to work around this is to use the fact that for sufficiently large $t_\infty$ the most unstable mode(s) are still identified accurately. Another set of impulse response simulations can then be run with these modes projected out in order to identify the more stable modes. The procedure can potentially be repeated until each unstable mode has been identified and projected out. In this case the method is clearly not {\it projection-free} anymore. It becomes more similar to the one suggested by Barbagallo {\it et al.} \cite{Barbagallo2009} and Ahuja {\it et al.} \cite{Ahuja2010}, although here the entire unstable subspace does not necessarily need to be projected out for the method to work. Additionally, it still avoids the need for an Arnoldi solver and theoretically still yields the same transformation $T$ as the projection-free method (and hence the unstable subspace is still balanced).

An alternative approach is inspired from the work of Morgans and Dowling \cite{Morgans2007} and Illingworth, Morgans and Rowley   \cite{Illingworth2011}: the unsatisfactory ROM obtained from the impulse response can be used as a first approximation to the system (with again the most unstable mode(s) accurately identified). Using this ROM, a controller aimed at suppressing the most unstable mode(s) can be designed and the resulting closed-loop impulse response can be used to deduce the corresponding open-loop dynamics and improve the initial ROM.

\subsubsection{Large systems and Gaussian quadrature}
In order to generate the balanced ROM, snapshots are typically recorded at regular time intervals. The matrices $X$ and $Z$ as defined in Eq.~(\ref{eq:xzapprox}) - which are used to approximate the Gramian integrals in Eq.~(\ref{eq:Wc}) and (\ref{eq:Wo}) - are formed by using Newton-Cotes quadrature weights of a selected order (e.g. trapezoidal, Simpson or Boole rule). However, the number of snapshots required can be significantly reduced by selecting a Gaussian quadrature rule instead (e.g. Gauss-Legendre quadrature), where the snapshots are not equally spaced in time.

Since the final simulation time is usually not known {\it a priori}, it may be preferable to use a ``composite'' quadrature rule. In this case each impulse response is divided into several time windows (of potentially different lengths). If there are $N_i$ snapshots in the $i$th time window, the integral of this window can then be evaluated independently from the rest of the impulse response, using a $(N_i)$-point Gaussian quadrature. With this piecewise integration method, it becomes straightforward to add an additional time window to increase the final simulation time. It is then also possible to optimize the distribution of the snapshots across the full simulation, for instance, if a higher-order quadrature is required in a specific time window. 

If the system dimension is so large that storing each snapshot is computationally demanding or if the number of snapshots required is so large that it results in an excessively expensive SVD (for three-dimensional flows for instance), then it may be necessary to use Gaussian quadrature to make the balanced POD method tractable.


\section{Application to a spatially developing unstable system \label{applitogle}}
\subsection{Linearized complex Ginzburg-Landau equation simulation setup}
In order to check that the application of the approximate balanced truncation method to unstable systems results in satisfactory ROMs for large-scale problems, we first apply the method to the linearized complex Ginzburg-Landau equation. This one-dimensional system exhibits spatially developing behavior and instabilities, which are representative of typical flow fields. The linearized complex Ginzburg-Landau equations (\ref{eq:gle}) and the corresponding adjoint equations (\ref{eq:agle}) are:
\begin{subequations}
\begin{align}
&\frac{\partial q}{\partial t}={\cal A}q=\left(-\nu \frac{\partial}{\partial x}+\gamma \frac{\partial^2}{\partial x^2} +\mu(x) \right)q\label{eq:gle},\\
&\frac{\partial q^+}{\partial t}={\cal A^+}q^+=\left(\nu^* \frac{\partial}{\partial x}+\gamma^* \frac{\partial^2}{\partial x^2} +\mu(x)^* \right)q^+\label{eq:agle},
\end{align}
\end{subequations}
where $\mu(x)=\mu_0-c_u^2+\mu_2x^2/2$. Note that now $x$ is the spatial variable, while $q$ is the system state. The complex parameters $\nu$ and $\gamma$ characterize the convection and diffusion/dispersion in the system respectively, while $\mu$ is related to the exponential growth of instabilities. $\mu_0$ can be seen as being similar to the Reynolds number in the Navier-Stokes equations, as it is used to determine the nature of the global stability of the system.  A large value of $\mu_2$ corresponds to a large degree of non-parallelism, while a small value of $\mu_2$ and a large value of $\nu$ result in a  strongly non-normal flow. Finally $c_u$ is the most unstable wavenumber in the flow-field. 

The results in this section were obtained by using the code developed by Bagheri \cite{Bagheri2009c} with a similar set of parameters to the supercritical (globally unstable) system considered in \cite{Bagheri2009c}, as summarized in Table \ref{paratable}. In order to demonstrate the method's ability to deal with several unstable modes however, $\mu_0$ was set to $0.57$ in order to obtain a two-dimensional antistable subspace. In this code, the forward problem uses a spectral Hermite collocation method, where state $q\in{\mathbb C^{n_x}}$ is evaluated at the roots of the $n_x$th Hermite polynomial $H_{n_x}(b x)$, and the parameter $b$ is chosen to obtain an accurate approximation of the continuous problem. The adjoint equations have been obtained from the discretized forward equations using a scaling matrix $M$, so that the energy of the discretized state, defined using the inner product $q^\dagger M q$, approximates the energy of the continuous state. Therefore, the adjoint state-space matrices are of the form ${(A^+,B^+,C^+)=(M^{-1}A^\dagger M,M^{-1}C^\dagger,B^\dagger M)}$, as opposed to simply ${(A^\dagger ,C^\dagger,B^\dagger )}$ (see the Appendix of \cite{Bagheri2009c} for more details regarding the discretization of the code).

In the present work, the system was set up as a single-input-single-output (SISO) system: the input (actuator) has a narrow Gaussian distribution, centered at $x_I$, the upstream limit of the region where instabilities are able to grow (``branch I'' in \cite{Bagheri2009c}). The output (sensor) is defined by the same Gaussian function, but centered at $x_{II}$, the downstream limit of the growth region (``branch II''). The discretized system has $220$ states, corresponding to a spatial extent of ${\left[-85,85\right]}$.

\begin{table}
\begin{center}
\begin{tabular}{lc}
Parameter&Value\\\hline
$n_x$&$220$\\
$\mu_0$&$0.57$\\
$\mu_2$&$-0.01$\\
$\nu$&$2+0.2i$\\
$\gamma$&$1-i$\\
$x_I$, $x_{II}$&$\pm 10.7$\\
$c_u$&$0.2$
\end{tabular}
\caption{Ginzburg-Landau equation simulation parameters\label{paratable}}
\end{center}
\end{table}

The Ginzburg-Landau equation has been often used to model fluid instabilities (reviews on this topic such as \cite{Huerre1990, Chomaz2005} frequently demonstrate important concepts with this one-dimensional model), and is a common test case for flow-control and model reduction  studies of convectively and globally unstable flows because its behavior is often representative of the Navier-Stokes equations but at a much lower computational cost (e.g.  \cite{Park1993, Lauga2003a, Cohen2005}). For more details about the Ginzburg-Landau equation and related studies, the reader is referred to \cite{Bagheri2009c}.

\subsection{Numerical results}
In this case, the cost of a simulation is low, so there was no need to use a Gaussian quadrature scheme suggested in Sec.~\ref{practical} and instead a Boole rule quadrature scheme was chosen. The number of snapshots used was chosen such that $t_\infty/N\approx0.05$. The balancing transformations and ROM were then obtained as described in Sec.~\ref{baltruncstable} and Sec.~\ref{newbaltrunc}. As there are only $220$ states in the full system, it was also possible to compute the transformations and ROMs based on the various procedures described in Sec.~\ref{baltruncunstable}. For clarity however, we choose to compare our method to the snapshot-based projection method \cite{Barbagallo2009, Ahuja2010} and what can be referred to as its ``exact'' equivalent \cite{Enns1985}, where the stable subsystem is balanced exactly.

\subsubsection{Comparison of the system's behavior with theory \label{theorycheck}}
As the conclusions of Sec.~\ref{newbaltrunc} are based on the theoretical limiting behavior of unstable systems, it is crucial to investigate the extent to which the key steps of the method yield the predicted results. In this section, we therefore compare several matrices obtained from the simulations with their theoretical counterparts. When the matrices considered here have a dual analog, which behaves in the same way, we choose to focus only on one of them for brevity.

\begin{figure}
\begin{center}
\includegraphics[width=0.5\columnwidth]{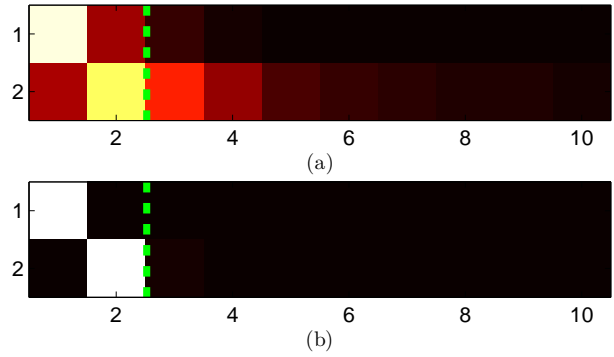}
\caption{(Color Online) First 10 columns of ${\rvert\tilde U_c^\dagger  U_c\lvert}$ for $t_\infty=20$ (a) and $t_\infty=80$ (b). The green dashed line separates the stable and antistable parts of the matrix. The color scale is from 0 (black) to 1 (white).\label{ucmatrices}}
\end{center}
\end{figure}

The first result that is checked here is part of the conclusion of {\it Proposition 1}, i.e. that $\lim_{t_\infty\to+\infty}u_{ci}= \tilde u_{ci}$, where ${\tilde u_{ci}=T_{ci-1}p_i\rVert T_{ci-1}p_i\lVert^{-1}}$ for unstable modes (there are two in this case), as stated in Eq.~(\ref{eq:prop1a}). We therefore plot the first 10 columns of ${\lvert\tilde U_c^\dagger U_c\rvert}$, with ${\tilde U_c=\left[\begin{array}{cc}\tilde u_{c1}&\tilde u_{c2}\end{array}\right]}$, which we expect to be equal to ${\left[\begin{array}{cc}I^{2\times 2}&0^{2\times 8}\end{array}\right]}$. Figure \ref{ucmatrices} shows that indeed, for large enough $t_\infty$, the first unstable singular vector of the controllability Gramian $u_{c1}$ points in the direction of $p_1$ and the second singular vector $u_{c2}$ points in the direction of the component of $p_2$ that is orthogonal to $p_1$. An analogous conclusion can be drawn for $U_o$ and Eq.~(\ref{eq:prop1b}).

\begin{figure}
\begin{center}
\includegraphics[width=0.5\columnwidth]{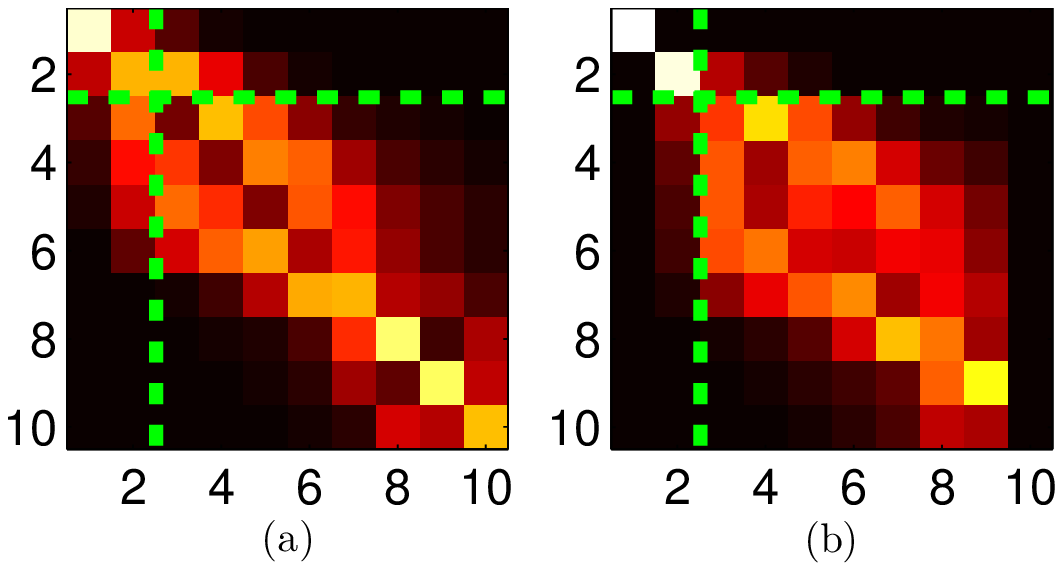}
\caption{(Color Online) Top left $(10\times10)$ elements of ${\rvert V_c^\dagger V\lvert}$ for ${t_\infty=20}$ (a) and $t_\infty=80$ (b). The green dashed lines separate the stable and antistable parts of the matrix. The color scale is from 0 (black) to 1 (white).\label{uvmatrices}}
\end{center}
\end{figure}

The next result of interest is an intermediate conclusion of {\it Proposition 2}: Eq.~(\ref{eq:hanksingvals}) states that the left and right unstable Hankel singular vectors tend to the right unstable singular vectors of $Z$ and $X$ respectively. We therefore expect the following to hold: 
\begin{eqnarray}
\lim_{t_\infty\to+\infty} V_c^\dagger V
\!=\!\lim_{t_\infty\to+\infty}\left[\begin{array}{cc}V_{ca}^\dagger V_{a}&V_{ca}^\dagger V_{s}\\V_{cs}^\dagger V_{a}&V_{cs}^\dagger V_s\end{array}\right]
\!=\!\left[\!\begin{array}{cc}I^{2\times 2}&0\\0&V_{cs}^\dagger V_s\end{array}\!\right]\!.\nonumber
\end{eqnarray}
Indeed for large enough $t_\infty$, Fig.~\ref{uvmatrices} shows that this is approximately true although at $t_\infty=80$, residuals still appear in the cross-terms $V_{ca}^\dagger V_{s}$ and $V_{cs}^\dagger V_{a}$.

\begin{figure}
\begin{center}
\includegraphics[width=0.5\columnwidth]{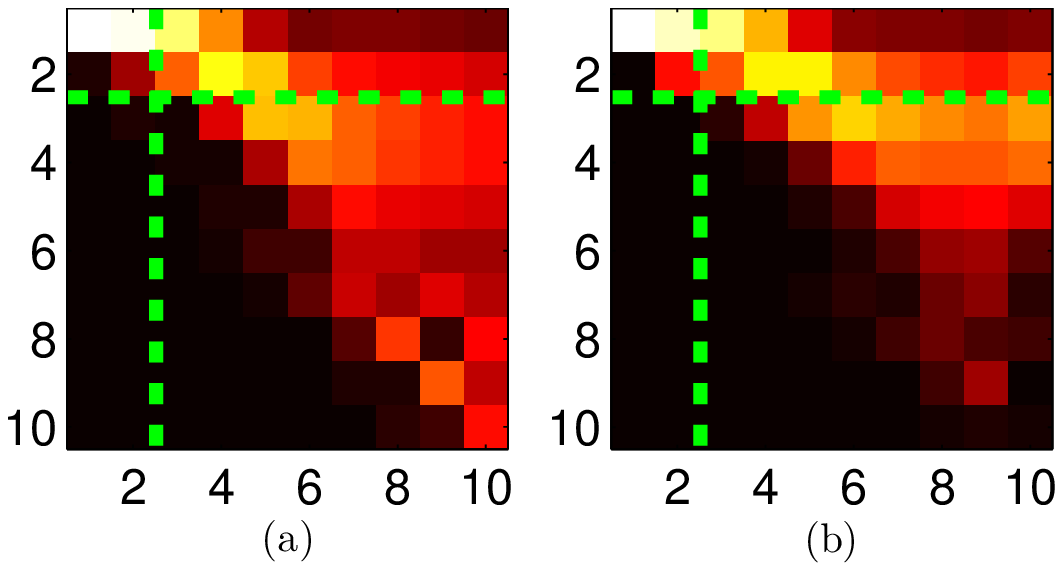}
\caption{(Color Online) Top left $(10\times10)$ elements of ${\rvert U_c^\dagger \hat T\lvert}$ for $t_\infty=20$ (a) and $t_\infty=80$ (b). The green dashed lines separate the stable and antistable parts of the matrix. The color scale is from 0 (black) to 1 (white).\label{tmatrices}}
\end{center}
\end{figure}
We now turn our attention to the transformation matrix, where we theoretically expect the direction of the unstable balanced modes to tend to that of the unstable left controllability singular vectors $U_{ca}$. In Fig.~\ref{tmatrices}, we plot $\lvert U_c^\dagger \hat T\rvert$, where $\hat T_{i}$ are the normalized balanced modes $\hat T_{i}=T_{i}\lVert\hat T_{i}\rVert^{-1}$ so each column is in the same direction as the corresponding column of $U_c^\dagger T$:
\begin{eqnarray}
\lim_{t_\infty\to+\infty}\!U_c^\dagger T\!
&=&\!\lim_{t_\infty\to+\infty}\Sigma_cV_c^\dagger V\Sigma^{-1/2},\nonumber\\
&=&\!\!\lim_{t_\infty\to+\infty}\!\left[\!\!\begin{array}{cc}\Sigma_{ca}V_{ca}^\dagger V_{a}\Sigma_{a}^{-1/2}\!&\Sigma_{ca}V_{ca}^\dagger V_{s}\Sigma_{s}^{-1/2}\\
\Sigma_{cs}V_{cs}^\dagger V_{a}\Sigma_{a}^{-1/2}\!&\Sigma_{cs}V_{cs}^\dagger V_s\Sigma_{s}^{-1/2}\end{array}\!\!\right]\!\!,\nonumber\\
&=&\left[\begin{array}{cc}\Sigma_{ca}\Sigma_{a}^{-1/2}&0\\0&U_{cs}^\dagger  T_s\end{array}\right].\nonumber
\end{eqnarray}
We therefore expect:
\begin{eqnarray}
\lim_{t_\infty\to+\infty} \!\!U_c^\dagger \hat T\!=\!
\lim_{t_\infty\to+\infty}\left[\begin{array}{cc}U_{ca}^\dagger \hat  T_a&U_{ca}^\dagger \hat  T_s\\U_{cs}^\dagger \hat  T_a&U_{cs}^\dagger \hat  T_s\end{array}\right]
\!=\!\left[\begin{array}{cc}I&0\\0&U_{cs}^\dagger \hat  T_s\end{array}\right]\!.\nonumber
\end{eqnarray}
However, as seen in Fig.~\ref{uvmatrices}, ${V_c^\dagger V}$ is not fully converged at $t_\infty=80$ and hence the only part of the expected behavior that clearly appears in Fig.~\ref{tmatrices} is ${\lvert U_{cs}^\dagger \hat T_a\rvert=0}$. By recalling that $U_{cs}$ spans the component of the stable subspace that is orthogonal to the antistable subspace, this can be interpreted as the fact that the unstable balanced modes span the same subspace as the unstable global modes in practice. However, it does {\it not} seem that the stable balanced modes are orthogonal to the unstable subspace at $t_\infty=80$ since ${\lvert U_{ca}^\dagger \hat T_s\rvert\ne0}$, or that each unstable balanced mode is in the direction of the corresponding controllability singular vector since ${\lvert U_{ca}^\dagger \hat T_a\rvert\ne I}$. These observations can be explained by considering the structure of ${\rvert U_c^\dagger \hat T\lvert}$ when $V_c^\dagger V$ is still converging and is instead of the form:
\begin{eqnarray}
V_c^\dagger V=\left[\begin{array}{cc}I+\epsilon_{aa}&\epsilon_{as}\\\epsilon_{sa}&U_{cs}^\dagger \hat  T_s\end{array}\right],
\end{eqnarray}
for some potentially small $\epsilon_{aa}$, $\epsilon_{as}$, and $\epsilon_{sa}$ matrices of the correct dimensions:
\begin{eqnarray}
U_c^\dagger T=\left[\begin{array}{cc}\Sigma_{ca}\left(I+\epsilon_{as}\right)\Sigma_{a}^{-1/2}&\Sigma_{ca}\epsilon_{as}\Sigma_{s}^{-1/2}\\
\Sigma_{cs}\epsilon_{sa}\Sigma_{a}^{-1/2}&U_{cs}^\dagger  T_s\end{array}\right].\nonumber
\end{eqnarray}
Given that ${\Sigma_{cs}}$ and ${\Sigma_{s}}$ converge to constants, while ${\lim_{t_\infty\to+\infty}\Sigma_{a}=+\infty}$ and ${\lim_{t_\infty\to+\infty}\Sigma_{ca}=+\infty}$, it is not surprising that ${\lim_{t_\infty\to+\infty}\Sigma_{cs}\epsilon_{sa}\Sigma_{a}^{-1/2}=0}$ and that unless all the elements of $\epsilon_{as}$ are exactly $0$, ${\lim_{t_\infty\to+\infty}\Sigma_{ca}\epsilon_{as}\Sigma_{s}^{-1/2}\ne0}$. Similarly, the off-diagonal elements of ${\Sigma_{ca}\left(I+\epsilon_{as}\right)\Sigma_{a}^{-1/2}}$ are of the form: ${\sigma_{cai}\epsilon_{aaij}\sigma_{aj}^{-1/2}}$ for some potentially small scalar ${\epsilon_{aaij}}$. For large $t_\infty$, if ${i>j}$ (below the diagonal), then $\sigma_{cai}\ll\sigma_{aj}^{1/2}$, whereas if ${i<j}$ (above the diagonal), we have ${\sigma_{cai}\gg\sigma_{aj}^{1/2}}$. This therefore explains the upper-triangular nature of ${\Sigma_{ca}\left(I+\epsilon_{as}\right)\Sigma_{a}^{-1/2}}$.

\begin{figure}
\begin{center}
\includegraphics[width=0.5\columnwidth]{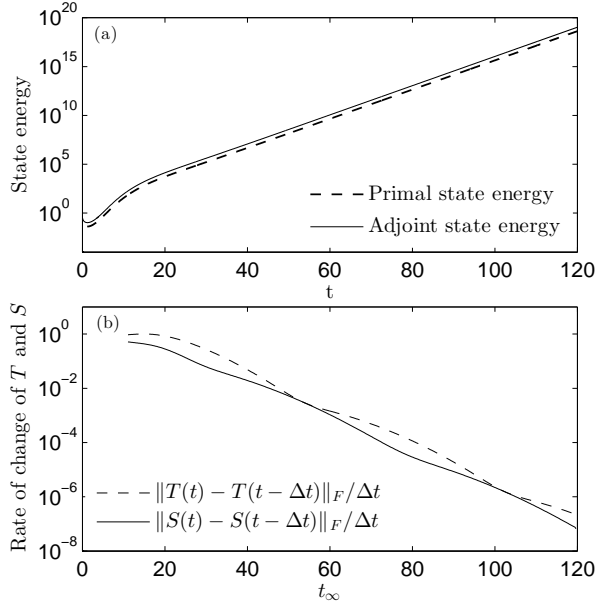}
\caption{Convergence of the balancing transformations $T$ and $S$. (a) Evolution of the energy of the forward impulse response (dashed line) and adjoint impulse response (solid line) states with respect to time. (b) Rate of change of the transformation matrices with respect to the final simulation time $t_\infty$ used to compute them, for the first $n_x$ columns of $T$ and rows of $S$ where $\Delta t=0.2$.
\label{TSchanges}}
\end{center}
\end{figure}

The behavior of the converging matrices considered up to this point can therefore readily be explained. We now wish to check that the transformation matrices converge as expected in practice. In order to do this, the transformation matrices $T$ and $S$ are computed for a range of $t_\infty$ values, and the rate of change in the transformations over a fixed time interval $\Delta t$: ${\lVert T(t_\infty)-T(t_\infty-\Delta t)\rVert_F/\Delta t}$ and ${\lVert S(t_\infty)-S(t_\infty-\Delta t)\rVert_F/\Delta t}$ are shown in Fig.~\ref{TSchanges}b ($\lVert\cdot\rVert_F$ is the Frobenius norm of a matrix). Clearly both matrices converge to constants for large $t_\infty$, as expected despite the exponential growth of the energy of the primal and adjoint states, shown in Fig.~\ref{TSchanges}a.

\begin{figure}
\begin{center}
\includegraphics[width=0.5\columnwidth]{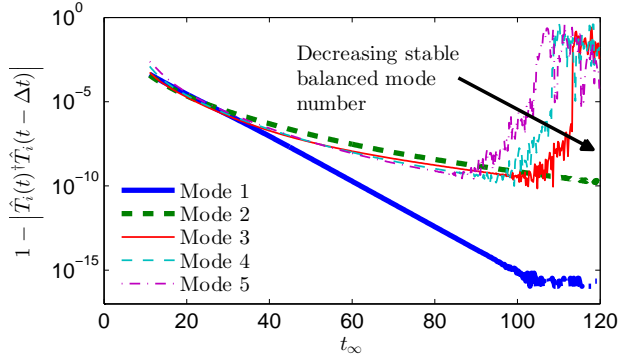}
\caption{(Color Online) Convergence of the balanced mode shapes, shown by plotting ${1-\lvert\hat T_i(t)^\dagger \hat T_i(t-\Delta t)\rvert}$ with respect to the final simulation time $t_\infty$ used to compute $T$, for the first 5 normalized balanced modes ${\hat T_i=T_i\rVert T_i\lVert^{-1}}$. Here $\Delta t=0.2$ and unstable modes are shown with thick lines.
\label{Tderivative}}
\end{center}
\end{figure}

It is important to note that although $T$ and $S$ converge, the convergence of the {\it direction} of the balanced modes also needs to be checked. We can therefore plot ${1-\rvert\hat T_i(t)^\dagger \hat T_i(t-\Delta t)\lvert}$ for each normalized balanced mode $\hat T_i$ to check that the direction of the balanced modes also converges. Figure \ref{Tderivative} shows that even for large $t_\infty$, the most unstable mode converges, while the more stable modes eventually become so negligible compared to the most unstable mode that they lose their accuracy. Not surprisingly, the less dynamically significant the mode, the lower the value of the maximum $t_\infty$ before it diverges. This behavior is expected as was mentioned in Sec.~\ref{practical} and the consequences of it for the quality of the ROM are further investigated in Sec.~\ref{ROManalysis} and in the next paragraph.

Finally, we check that the Gramians are indeed balanced by the transformation. By definition, Gramians computed using a given set of snapshots are exactly balanced by transformations $T$ and $S$, which were computed using the same set of snapshots. However, we wish to investigate how well the ``converged'' transformation balances Gramians computed using another set of snapshots. In order to do this, recall from Eq.~(\ref{eq:diagWo}) that the transformed observability Gramian is ${\Sigma_{(11)}^{-1/2}\left(V_{(11)}^\dagger V_{(21)}\right)\Sigma_{(21)}^2 \left(V_{(11)}^\dagger V_{(21)}\right)^\dagger\Sigma_{(11)}^{-1/2}}$. The only term that is potentially non-diagonal, $V_{(11)}^\dagger V_{(21)}$, therefore provides a good indication of how well-balanced the Gramians are, as we expect it to tend to the identity matrix. Similarly, we could investigate how close $U_{(11)}^\dagger U_{(12)}$ is to the identity matrix to investigate how well-balanced the controllability matrix is. 

We therefore choose to compute ${\lVert I-\lvert V_{(11)}^\dagger V_{(21)}\rvert\rVert_{F}}$, where the transformations have been computed using a set of snapshots corresponding to $t_\infty=t_1=20$, $40$, and $80$ time units and applied to Gramians, which were computed for values of $t_\infty=t_2$ ranging from 0 to 120. We only consider the top left $(4\times 4)$, $(8\times 8)$, and $(12\times 12)$ elements of $V_{(11)}^\dagger V_{(21)}$, to evaluate how balanced the resulting $4$th, $8$th, and $12$th-order ROMs would be. 

\begin{figure*}
\includegraphics[width=\textwidth]{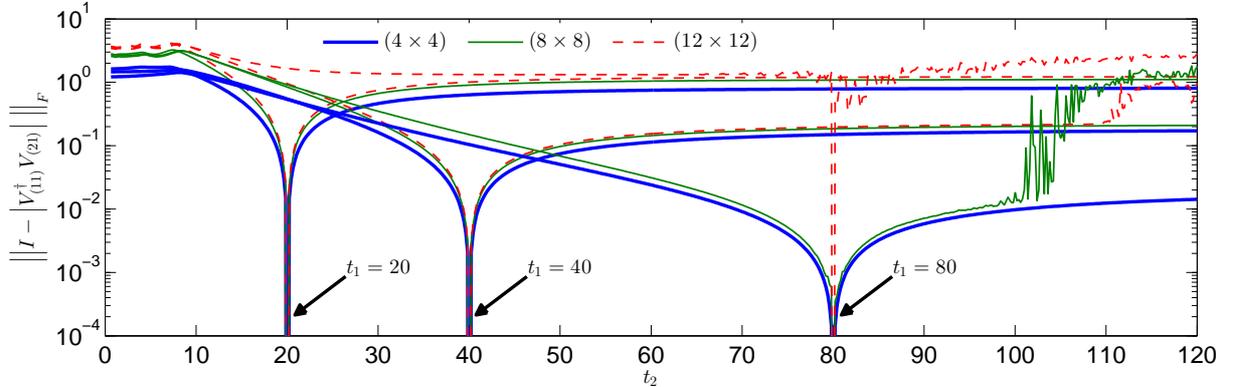}
\caption{(Color Online) Balancing error of transformed Gramians computed from a simulation with final simulation time $t_\infty=t_2$ and balanced using transformation matrices calculated from simulations with $t_\infty=t_1$. The error is quantified by the top left $(4\times 4)$ (thick), $(8\times 8)$ (thin), and $(12\times 12)$ (dashed) elements of $\lVert I-\lvert V_{(11)}^\dagger V_{(21)}\rvert\rVert_{F}$ for ${t_\infty=t_1=20}$, $40$, and $80$. The value of $t_1$ from each curve can be identified by the zero error values at $t_2=t_1$.
\label{VHV}}
\end{figure*}

Focusing first on the influence of $t_1$, Fig.~\ref{VHV} shows that in general, computing the matrix with a longer $t_1$ improves the balancing for all three orders of the ROM (for all except highest values of $t_2$ and close to $t_1=t_2$), as expected: theoretically, the higher the value of $t_1$, the more converged the transformation matrices. Specifically, the three curves corresponding to $t_1=20$ have a large error regardless of the ROM order, as the transformation matrices are not converged. On the other hand, if $t_1$ is too large, the set of snapshots used to define the transformation does not allow an accurate identification of dynamically less significant modes. Therefore Gramians (and ROMs) which include these modes will not be properly balanced regardless of $t_2$. In Fig.~\ref{VHV} therefore, the curve corresponding to a 12-mode Gramian with $t_1=80$ always has a large balancing error.

Considering now the effect of $t_2$, the quality of the balancing increases as $t_2$ approaches $t_1$. At $t_2=t_1$ the Gramians are exactly balanced by definition as mentioned above. For $t_2>t_1$, the error settles to a constant value, until eventually, a sharp degradation appears for the highest values of $t_2$, for instance at $t_2\approx100$ for the $t_1=80$, 8-mode ROM. 
This behavior can be explained by the fact that the accuracy of the Gramians, $XX^\dagger$ and $ZZ^\dagger$, as opposed to the transformations, decreases if $t_2$ is too large, as they are also approximated using state snapshots.

In this section, we have shown that in practice, we cannot allow the system to fully converge by letting $t_\infty\to+\infty$, as this results in the loss of crucial information about the dynamically significant stable behavior of the system. However, the balanced modes do converge up to a critical $t_\infty$ value and result in approximately balanced transformed Gramians. We therefore now turn our attention to the physical problem of the performance of reduced-order models obtained with this method, and compare them with ROMs obtained with existing methods.

\subsubsection{Analysis of the reduced-order models \label{ROManalysis}}
In this article, the ${\cal L}_\infty$ norm of the difference between the full-order transfer function and the reduced-order transfer function, defined as ${\sup_{\omega\in\mathbb{R}\cup \infty}{\lvert G(i\omega)-G_r(i\omega)\rvert}}$ for SISO systems is used as a measure of ROM quality. In practice, we approximate this by: ${\max_{\omega\in\left[-\omega_{\infty},\omega_{\infty}\right]} {\lvert G(i\omega)-G_r(i\omega)\rvert}}$ for ${\omega_\infty\to+\infty}$. Note that this transfer function norm is different both from the ${\cal H}_\infty$ error defined as ${\sup_{s:\text{Re}(s)>0}{\lvert G(s)-G_r(s)\rvert}}$, which is infinite for unstable systems, and from the time-domain ${\cal L}_\infty(I_t)$ norm usually used for signals $x(t)$ and defined over some interval $I_t$ as $\text{ess sup}_{t\in I_t}\lvert x(t)\rvert$. The ${\cal L}_\infty$ error norm normalized by ${\cal L}_\infty$ norm of the full-order system will be referred to as the ``ROM error'' in the following paragraphs, i.e. ${\lVert G-G_r\rVert_\infty}/\lVert G\rVert_\infty$. Potentially unstable systems can alternatively be compared using the $\delta_\nu$-gap metric \cite{Vinnicombe1993}, which can be interpreted as the ``distance'' between two plants from the point of view of their behavior in a feedback setting. However for the purposes of the present analysis, the ${\cal L}_\infty$ norm is an adequate measure of the difference between the two systems (in this case it represents the maximum distance between the loci of $G(i\omega)$ and $G_r(i\omega)$), as our focus is on model reduction as opposed to controller design in this article.

\begin{figure*}
\includegraphics[width=\textwidth]{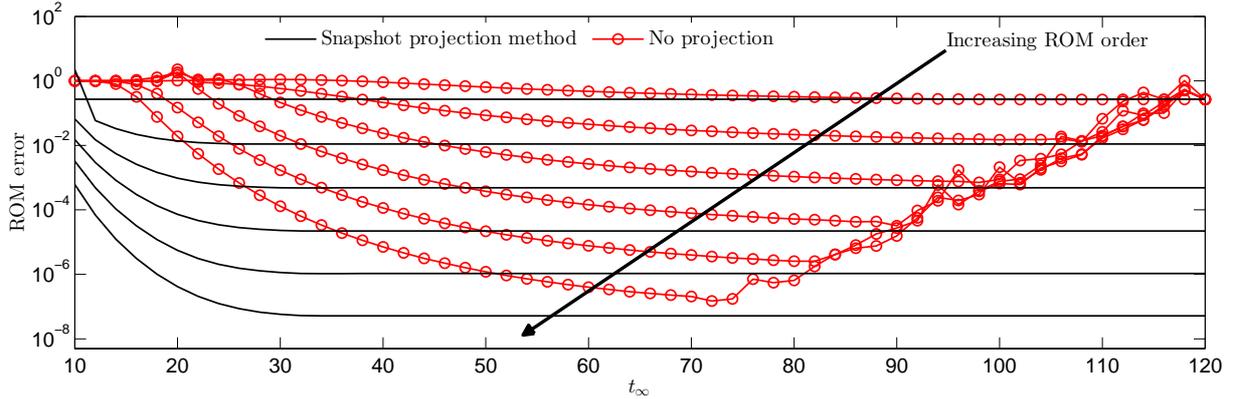}
\caption{(Color Online) ROM error as a function of the final simulation time $t_\infty$. Solid lines: snapshot projection method \cite{Barbagallo2009, Ahuja2010}. Lines with circles: projection-free method. For both methods, the different lines correspond to ROMs with 2, 4, 6, 8, 10 and 12 states, from top to bottom.\label{tmaxconv}}
\end{figure*}

\begin{figure}
\begin{center}
\includegraphics[width=0.5\columnwidth]{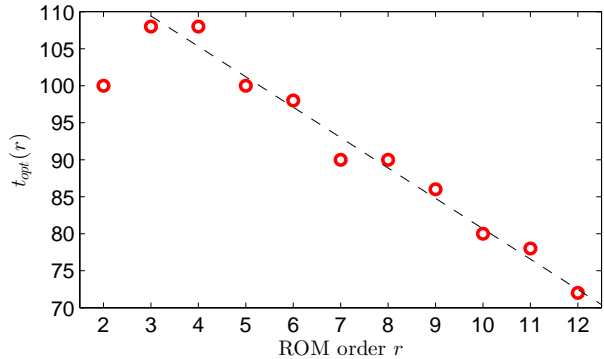}
\caption{(Color Online) Estimated optimal final simulation time $t_{opt}(r)$ for different ROM orders $r$ (symbols) and linear tend line plotted for $r>2$. \label{Opt_tinf}}
\end{center}
\end{figure}

Given the restrictions on the final simulation time mentioned in Sec.~\ref{practical} and Sec.~\ref{theorycheck}, we first investigate the quality of ROMs obtained with different $t_\infty$ in Fig.~\ref{tmaxconv}. The behavior described in Sec.~\ref{practical} is apparent: as the simulation time increases the ROM error initially decreases. The lower bound on $t_\infty$ related to the decay of the stable modes is illustrated with the black lines with no markers, corresponding to ROMs of different orders, but computed with the projection method of \cite{Barbagallo2009, Ahuja2010}. The second lower bound on $t_\infty$, corresponding to the convergence of the unstable modes as they begin to dominate the impulse response appears to be slower, as shown by the lines with markers, corresponding to projection-free ROMs. Now, if $t_\infty$  is sufficiently large for the less dynamically significant information to be lost, increasing the order of the ROM does not reduce the error any further. Additionally, each ROM order has a different optimal final simulation time: the error is minimized just before the unstable modes start to dominate the response enough to sharply increase the error. In the following paragraphs, an estimated value $t_{opt}(r)$ for the optimal $t_\infty$ corresponding to each ROM order $r$ is used. These values are shown in Fig.~\ref{Opt_tinf}, where it can be observed for this particular case that $t_{opt}(r)$ seems to be decreasing roughly linearly as the model order is increased for $r>2$.

\begin{figure*}
\includegraphics[width=\textwidth]{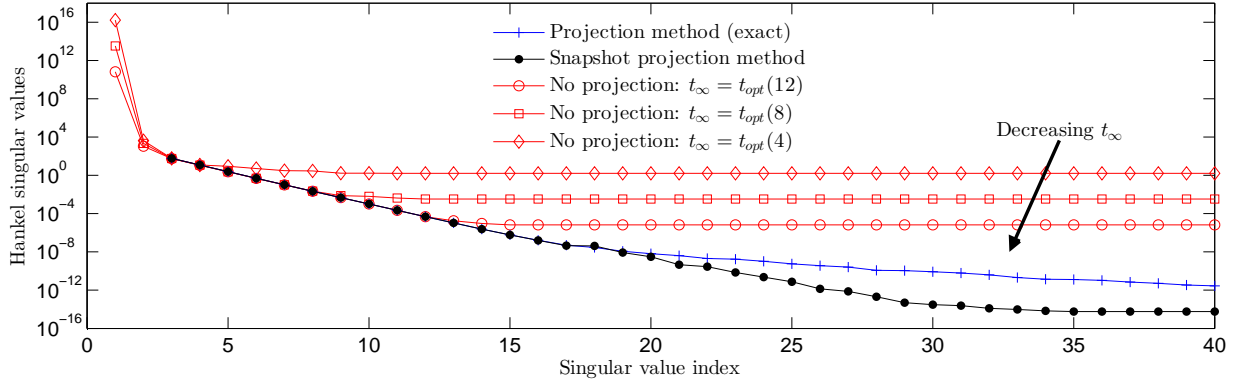}
\caption{(Color Online) Hankel singular values of ROMs obtained with the exact projection method \cite{Enns1985} (crosses), the snapshot-based projection method \cite{Barbagallo2009, Ahuja2010} (filled circles), and the projection-free method (open symbols) obtained with $t_\infty= t_{opt}(r)$ for ROM orders of 4 (diamonds), 8 (squares), and 12 (open circles), from top to bottom. Note that the unstable singular values are not plotted for the first two methods as they are infinite by definition.\label{HSVs}}
\end{figure*}

The existence of an optimal final simulation time can be further illustrated by plotting the singular values of the Hankel matrix (Fig.~\ref{HSVs}) for different values of $t_\infty$. Here, the HSVs corresponding to 4th, 8th and 12th-order ROMs, obtained with the exact and snapshot projection methods as well as the projection-free method (using estimated optimal $t_\infty$ values) are plotted. As $t_\infty$ is increased, in the projection-free case, the HSVs corresponding to unstable modes grow, while the rest of the singular values tend towards their respective converged values. This explains the increasing precision of the ROMs as the final simulation time is increased, like in stable systems. On the other hand, the flat region of each curve, where the singular values effectively have no dynamical significance, incorporates an increasing number of states as $t_\infty$ grows. This again corresponds to the less significant information disappearing due to the unstable modes becoming too large, and can be used as a way to choose the appropriate ROM order, given an HSV distribution. For instance, with $t_\infty=t_{opt}(4)=108$ (top line with diamonds in Fig.~\ref{HSVs}), roughly 4 singular values are not in the flat region of the curve. As shown in Fig.~\ref{tmaxconv}, the error will start growing if $t_\infty$ is increased further with $r=4$. Conversely, increasing the ROM order to anything higher than 4 will not reduce the ROM error significantly when $t_\infty=t_{opt}(4)$.

\begin{figure}
\begin{center}
\includegraphics[width=0.5\columnwidth]{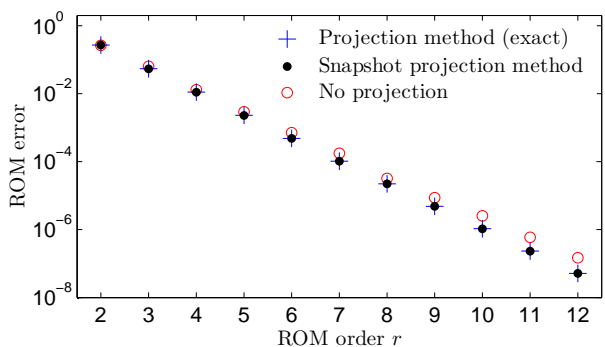}
\caption{(Color Online) ROM error as a function of the ROM order, obtained with $t_\infty= t_{opt}(r)$ for the projection-free method. Crosses: exact projection method  \cite{Enns1985}. Filled circles: snapshot-based projection method \cite{Barbagallo2009, Ahuja2010}. Open circles: projection-free method. \label{Error_ns}}
\end{center}
\end{figure}

In Fig.~\ref{Error_ns}, the error obtained with $t_{opt}(r)$ is plotted as a function of the ROM order, and compared to the performance of the snapshot \cite{Barbagallo2009, Ahuja2010} and exact \cite{Enns1985} projection methods. Clearly, the two projection approaches yield ROMs of the same quality. On the other hand, the projection-free method ROM error is of comparable order of magnitude, with the difference in the error increasing slightly with the ROM order. Note however that this difference is not significant: the error of a 12-state projection-free ROM is smaller than that of an 11-state ROM computed with the projection methods.

\begin{figure}
\begin{center}
\includegraphics[width=0.5\columnwidth]{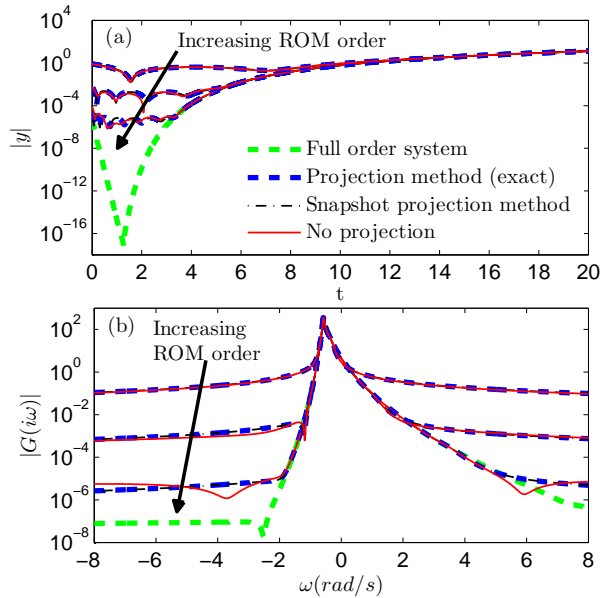}
\caption{(Color Online) Impulse response (a) and transfer function gain (b) of the full system (thick green dashed line) and the ROMs obtained with the exact projection method \cite{Enns1985} (thick blue dashed line), the snapshot-based projection method \cite{Barbagallo2009, Ahuja2010} (thin black dash-dotted line), and the projection-free method (thin red solid line). ROM orders: 4, 8, and 12, top to bottom for both the impulse responses and the transfer functions.\label{TFandIR}}
\end{center}
\end{figure}

For control design purposes, it is crucial that the ROMs are able to reproduce the full system's impulse response and transfer function. In Fig.~\ref{TFandIR}, the impulse responses (a) and the transfer function gains (b) of the three methods are compared for different ROM orders. The three methods yield similar results for the three ROM orders considered. As the ROM order in increased in all three cases, the initial transient of the impulse response is estimated with increasing precision. Similarly, the high-frequency gain is better modeled with high ROM orders in the transfer function.


\section{Application to a two-dimensional unstable flow-field \label{applitons}}
The previous sections in this article focused on demonstrating the basis for using the method from a theoretical point of view and by comparing its performance to that of existing methods, when applied to the linearized complex Ginzburg-Landau equation. In this section, the applicability of the method in a more realistic scenario, i.e. for large-scale unstable flow-fields is investigated. The chosen test case is the flow over a cylinder at Reynolds number ${Re=U_\infty D/\nu=100}$, where $U_\infty$ is the incoming flow velocity, $D$ is the cylinder diameter, and $\nu$ is the kinematic viscosity.

The forward direct numerical  simulations run here are based on the immersed boundary projection method code developed by Taira  and Colonius \cite{Taira2007, Colonius2008}. The two-dimensional, viscous, incompressible Navier-Stokes equations expressed in vorticity form are discretized on a second-order accurate finite-volume staggered grid. A multi-grid algorithm is used, where the problem is solved on a series of nested identical Cartesian uniform grids, each of twice the physical extent and half the resolution of the previous one. A second-order Adams-Bashforth scheme is used to discretize the advection term(s) and a second-order Crank-Nicolson scheme is used for the other (linear) terms. For more information, the reader is referred to  \cite{Taira2007, Colonius2008}. The linearized equations are therefore a discretization of:
\begin{eqnarray}
\begin{cases}
   \dot\omega=\nabla\!\times\!\left(u_0\!\times\!\omega\right)+
   \nabla\!\times\!\left(u\!\times\!\omega_0\right)\\
   \qquad-Re^{-1}\nabla\!\times\!\left(\nabla\!\times\!\omega\right)
   +\nabla\!\times\! f_{IB}+\nabla\!\times\! f,\label{eq:lns}\\
   u_B\!\!=0,
\end{cases}
\end{eqnarray}
where $\omega$ refers to vorticity, $u$ to velocity, $_0$ to base flow quantities, $f_{IB}$ to immersed boundary forcing that imposes the no-slip condition on the cylinder surface, i.e. the second equation in Eq.~(\ref{eq:lns}), where $u_B$ is the velocity on the body surface. The spatially distributed forcing function $f$ is used to model the input (more details below). 

The corresponding {\it continuous} adjoint equations can be shown to be:
\begin{eqnarray}
\begin{cases}
   -\dot\omega^+=\nabla\!\times\!\left(\omega_0\!\times\! u^+\right)-
   \nabla^2\left(u^+\!\times\! u_0\right)\\
   \qquad\quad-Re^{-1}\nabla\!\times\!\left(\nabla\!\times\!\omega^+\right)+
   \nabla\!\times\! f_{IB}^++\nabla\!\times\! f^+,\label{eq:ans}\\
   u_B^+=0,
\end{cases}
\end{eqnarray}
where $^+$ refers to adjoint quantities and $f^+$ forces the adjoint simulation at the sensor location (output location of the forward simulation), as opposed to $f$, which forces the forward simulation at the actuator (input) location.  In this code however, the adjoint equations were obtained from the spatially discrete but temporally continuous linearized equations, where the time-marching scheme is self-adjoint in the present case since the base flow is the steady unstable equilibrium of the problem. Note that the same nested-grid method was used in both the forward and adjoint problem and this procedure is not self-adjoint so the adjoint equations we solve are technically not the exact discrete adjoint of the forward problem. The solver used for Eq.~(\ref{eq:lns}) and (\ref{eq:ans}) is similar for both sets of equations, except for the advection terms and the fact that Eq.~(\ref{eq:ans}) runs backwards in time.

\begin{figure}
\begin{center}
\includegraphics[width=0.5\columnwidth]{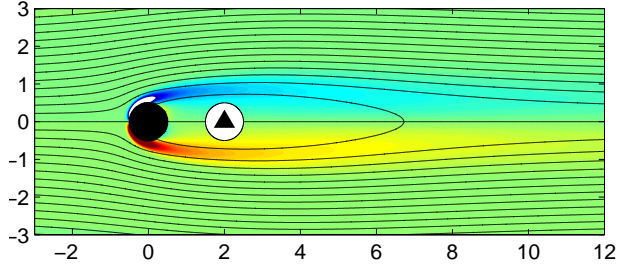}
\caption{(Color Online) Unstable base flow over a cylinder at $Re=100$, showing streamlines and contours of vorticity. The white circle represents the disk of diameter 1 where the input (a vertical body force) is applied and the black triangle shows the location of the output, a sensor measuring the vertical velocity at (2,0).\label{Baseflow}}
\end{center}
\end{figure}

The same grid as the one in \cite{Tu2012} was used here, where 3 nested grids were used, and where the smallest grid has dimensions $[-15,15]\times[-5,5]$ and the widest one ${[-60,60]\times[-20,20]}$. Each grid has ${1500\times500}$ square cells of length and width $\delta=0.02$. The chosen input-output setup is also almost identical to the one in \cite{Tu2012}: the flow is forced using a disk-shaped {\it vertical} body force of radius 1 (white disk in Fig.~\ref{Baseflow}), centered at ${(x,y)=(2,0)}$ in the wake. The sensor measures the {\it vertical} velocity, also at ${(x,y)=(2,0)}$ (black triangle in Fig.~\ref{Baseflow}). The base flow, shown in Fig.~\ref{Baseflow}, was computed using Selective Frequency Damping (SFD) \cite{Akervik2006, Jordi2014}. The SFD parameters values chosen here were ${\chi=0.4391}$ and ${\Delta=3.1974}$, as suggested by \cite{Jordi2014} for the cylinder flow at ${Re=100}$.

\begin{figure}
\begin{center}
\includegraphics[width=0.5\columnwidth]{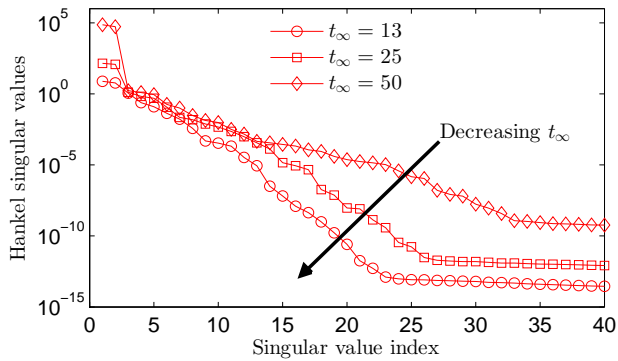}
\caption{(Color Online) Hankel singular values of ROMs obtained with the projection-free method with $t_\infty= 13$ (diamonds), $t_\infty= 25$ (squares), and $t_\infty= 50$ (open circles), from top to bottom.\label{HSVcylinder}}
\end{center}
\end{figure}

In order to compute the reduced-order model, both the forward and the adjoint impulse responses were run for a long time (300 time units). Snapshots were stored every 0.2 time units and a Boole quadrature rule was used to scale the snapshots. Once the snapshots are stored, one can compute the Hankel singular values of the ROMs for different values of $t_\infty$, as shown in Fig.~\ref{HSVcylinder}. The general behavior is similar to the one in Sec.~\ref{applitogle}, but in this case the HSV distribution is more complex than in Fig.~\ref{HSVs}. Furthermore, both the transients and the long-term response of the system are of interest, it is less clear how one should quantify the ROM error. Nevertheless, since the cost of computing the SVD of the Hankel matrix is relatively small, one can find a compromise between ROM order and quality by trial and error by comparing the model impulse response to the one from the sensor.

\begin{figure}
\begin{center}
\includegraphics[width=0.5\columnwidth]{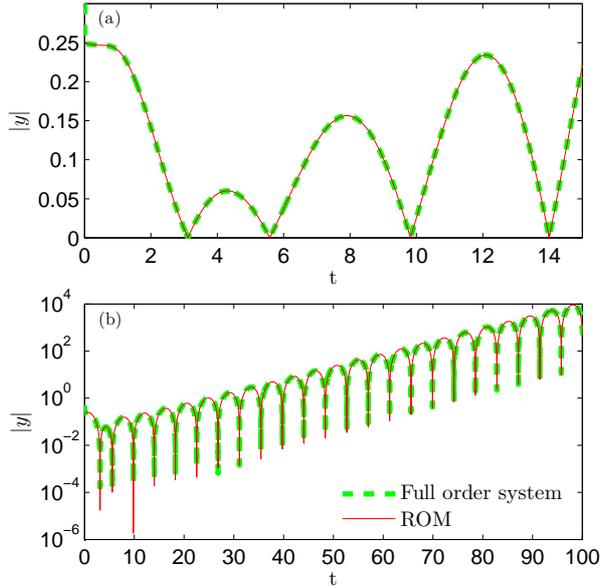}
\caption{(Color Online) Impulse response of the full cylinder flow (thick dashed line), compared to the 12th-order ROM obtained with the projection-free method (thin solid line). The agreement between the two impulse responses is excellent, both for the initial transients shown in (a) and long term unstable response shown in (b).\label{IRcylinder}}
\end{center}
\end{figure}

In this case, using only 125 snapshots, corresponding to a final simulation time of 25 time units, a 12th-order model was obtained, whose impulse response is compared to the full system's in Fig.~\ref{IRcylinder}. It is clear that the method successfully reproduces the impulse response in this case, both for times that are much longer than the final simulation time used to compute the model and for the initial transients. 

Note that the instability here is due to an unstable complex-conjugate pair of eigenvalues, i.e. the special case considered from a theoretical point of view in Appendix~\ref{unstableccpair}. This therefore demonstrates that the technique is also readily applicable in practice for this ubiquitous scenario. The HSVs corresponding to the two unstable complex conjugate modes clearly appear in Fig.~\ref{HSVcylinder} as the first two HSVs, which grow to large values as $t_\infty$ is increased, and are always of the same order of magnitude.

Another important point is that at $t=25$, the output is $\rvert y(t)\lvert \approx1$, which is only about 4 times larger than the first transient peak. In other words, we obtained a successful ROM of this unstable linear system without requiring an excessively long final simulation time or strong domination of the unstable modes. This suggests that with a sufficiently small impulse at the input, the method might be directly applicable to the initial linear response of some nonlinear systems.


\section{Conclusions \label{ccl}}
In this article, we have shown that applying the snapshot-based, projection-free, approximate balanced truncation method to unstable systems theoretically yields a converged transformation that balances the Gramians for all $t_\infty\to+\infty$, including the unstable subspace. For stable systems, this method has been popular in recent years as it is  straightforward to implement and scales well for large systems. The fact that it can be used without any modifications for unstable systems will allow the computation of BPOD-based ROMs and controllers of even three-dimensional unstable flow-fields, without the need to identify or project-out any global modes. 

Benchmarking the method using a one-dimensional, spatially developing, unstable system showed that it is not required for the system to converge to its theoretical limiting state for this approach to yield ROMs of similar quality to the ones obtained with existing methods \cite{Barbagallo2009, Ahuja2010, Enns1985}. In fact this limiting behavior cannot be reached exactly in practice due to finite-precision arithmetic: if the final simulation time becomes excessively large, only the unbounded growth of the most unstable modes will be conserved and the initial transient information from the impulse response will be lost. It was found that there is an optimal final simulation time value, which is a function of the desired ROM order, and methods to estimate this optimal value were discussed.

In cases where the unmodified method yields unacceptable ROMs, or where the balanced modes must be identified more accurately, a method was outlined to circumvent the final time restriction. This method has a slightly increased cost as it requires projecting out some of the system's unstable balanced modes, but their identification still {\it does not} require an Arnoldi solver. 
Additionally, if the system dimension is so large that storing a large number of snapshots is challenging, it was suggested that a (piecewise) Gaussian quadrature method (with varying sampling intervals) could be used when storing the snapshots, as opposed to a traditional Newton-Cotes approach with a constant sampling rate, as this is a simple way to significantly reduce the number of snapshots required to obtain a ROM of a given accuracy.

Finally, the method was applied to a more realistic large-scale system: the supercritical (linearized) flow over a circular cylinder at $Re=100$, and an excellent match was obtained with a 12th-order model. Only a short simulation and a small number of snapshots were required to obtain the ROM, where the final output amplitude was similar to the initial transients, suggesting that it may be possible to apply the method to the initial linear growth of nonlinear unstable flows.

The standard approximate balanced truncation algorithm was therefore shown to be applicable to unstable systems both from a theoretical and a practical point of view. The implications of this are twofold. First, the simple numerical setup required to compute the snapshot-based balanced truncation of stable systems can be safely applied without any modifications to unstable systems. Second, the computation of balanced reduced-order models of unstable systems  that are so large that other existing methods are not tractable is made possible with this approach.


\section{Acknowledgments}
This research is funded by the EPSRC. The authors would also like to thank Shervin Bagheri for allowing us to use the Ginzburg-Landau Equation {\sc matlab} code, on which Sec.~\ref{applitogle} is based, as well as Tim Colonius for the Immersed-Boundary Fractional Step code used in Sec.~\ref{applitons}.


\appendix
\section{Projection-free approximate balanced truncation of unstable systems: special cases \label{specialcases}}
In Sec.~\ref{newbaltrunc}, it was assumed that the growth rates of the different eigenmodes of the flow field were distinct, so that one eigenmode would eventually dominate the impulse response. The purpose of this section is to show that the proof extends in a similar way when some of the eigenmodes have the same growth rate. The general case where the $n$th to $(n+m)$th eigenvalues have the same growth rate: ${\text{Re}(\lambda_n)=\text{Re}(\lambda_{n+1})=\ldots=\text{Re}(\lambda_{n+m})=\alpha}$, but the imaginary parts ${\text{Im}(\lambda_n)=\omega_n}$ differ: $\omega_n\neq\omega_{n+1}\neq\ldots\neq\omega_{n+m}$ is considered first and the special cases of repeated eigenvalues and complex conjugate eigenvalue pairs are then treated separately.

\subsection{General case \label{appxgeneral}}
\subsubsection{Controllability and observability singular vectors}
Let us assume that the $(n-1)$ most unstable modes are projected out as in Sec.~\ref{newbaltrunc}. Let us also write ${\Omega=\text{diag}(\left[\begin{array}{ccc}\omega_n&\ldots&\omega_{n+m}\end{array}\right])}$. The impulse response state ${x(t)=e^{At}B}$ becomes:
\begin{eqnarray}
&&\lim_{t\to+\infty}T_{cn-1}x(t)\!=\!T_{cn-1}\!\left[\!\!\begin{array}{ccc}p_n&\!\!\ldots\!\!&p_{n+m}\end{array}\!\!\right]e^{i\Omega t}\!\left[\!\!\begin{array}{c}q^\dagger _{n}\\\vdots\\q^\dagger _{n+m}\end{array}\!\!\right]\!Be^{\alpha t},\nonumber\\
&&\Rightarrow\lim_{t_\infty\to+\infty}T_{cn-1}X=T_{cn-1}\left[\!\!\begin{array}{ccc}p_n&\ldots&p_{n+m}\end{array}\!\!\right]
               \left[\!\!\begin{array}{c}\beta^\dagger _{n}\\\vdots\\\beta^\dagger _{n+m}\end{array}\!\!\right]\nonumber\\
&&=\left[\begin{array}{ccc}u_{cn}&\ldots&u_{cn+m}\end{array}\right]
               \left[\begin{array}{ccc}\sigma_{cn}&&\\&\ddots&\\&&\sigma_{cn+m}\end{array}\right]
               \left[\begin{array}{c}v_{cn}^\dagger \\\vdots\\v_{cn+m}^{\dagger }\end{array}\right]\nonumber.
\end{eqnarray}
The left and right singular vectors of ${T_{cn-1}X}$ thus tend to the subspaces spanned by $T_{cn-1}\!\left[\!\!\begin{array}{ccc}p_n&\!\!\ldots\!\!&p_{n+m}\end{array}\!\!\right]$ and ${\left[\begin{array}{ccc}\beta_{n}&\ldots&\beta_{n+m}\end{array}\right]}$ respectively. $T_{on-1}Z$ behaves in an analogous way and hence its left and right singular vectors tend to the subspaces spanned by ${T_{on-1}\left[\begin{array}{ccc}q_n&\ldots&q_{n+m}\end{array}\right]}$ and ${\left[\begin{array}{ccc}\xi_{n}&\ldots&\xi_{n+m}\end{array}\right]}$ respectively. However ${\left[\begin{array}{ccc}u_{cn}&\ldots&u_{cn+m}\end{array}\right]}$ and ${\left[\begin{array}{ccc}u_{on}&\ldots&u_{on+m}\end{array}\right]}$ do not generally converge to constants due to the oscillatory term ${e^{i\Omega t}}$.

\subsubsection{Hankel matrix}
Next, as in Sec.~\ref{newbaltrunc}, we consider the projected Hankel matrix:
\begin{eqnarray}
&&\lim_{t_\infty\to+\infty}T_{ln-1}M_HT_{rn-1}=\lim_{t_\infty\to+\infty}T_{ln-1}Z^\dagger XT_{rn-1}\nonumber\\
&&=\left[\begin{array}{ccc}u_{n}&\ldots&u_{n+m}\end{array}\right]
               \left[\begin{array}{ccc}\sigma_{n}&&\\&\ddots&\\&&\sigma_{n+m}\end{array}\right]
               \left[\begin{array}{c}v_{n}^\dagger \\\vdots\\v_{n+m}^{\dagger }\end{array}\right],\nonumber\\
&&=\left[\begin{array}{ccc}v_{on}&\ldots&v_{on+m}\end{array}\right]M
               \left[\begin{array}{c}v_{cn}^\dagger \\\vdots\\v_{cn+m}^{\dagger }\end{array}\right],\nonumber\\
&&\text{where:}\nonumber\\
&&M\!=\!
               \left[\!\begin{array}{c}\sigma_{on}u_{on}^\dagger \\\vdots\\\sigma_{on+m}u_{on+m}^{\dagger }\end{array}\!\right]\!\!
               \left[\!\!\begin{array}{ccc}u_{cn}\sigma_{cn}&\!\!\ldots\!\!&u_{cn+m}\sigma_{cn+m}\end{array}\!\!\right]\label{eq:Mmatrix}\!\!.
\end{eqnarray}
The left and right Hankel singular vectors therefore tend to the subspaces spanned by $\left[\!\begin{array}{ccc}v_{on}&\!\ldots\!&v_{on+m}\end{array}\!\right]$ 
and ${\left[\begin{array}{ccc}v_{cn}&\ldots&v_{cn+m}\end{array}\right]}$ respectively, which are already (individually) orthonormal bases by definition. We thus only need unitary matrices to identify the Hankel singular vectors and values, which we can get from the SVD of $M$:
\begin{eqnarray}
\label{eq:Msvd}
M&=&R_o\left[\begin{array}{ccc}\sigma_{n}&&\\&\ddots&\\&&\sigma_{n+m}\end{array}\right]R_c^\dagger \,\,.
\end{eqnarray}
Therefore in general:
\begin{eqnarray}
\begin{array}{c}
\left[\begin{array}{ccc}u_{n}&\ldots&u_{n+m}\end{array}\right]
=\left[\begin{array}{ccc}v_{on}&\ldots&v_{on+m}\end{array}\right]R_o,\\\\
\left[\begin{array}{ccc}v_{n}&\ldots&v_{n+m}\end{array}\right]
=\left[\begin{array}{ccc}v_{cn}&\ldots&v_{cn+m}\end{array}\right]R_c,
\end{array}
\label{eq:repMHsv}
\end{eqnarray}
where $R_c$ and $R_o$ are not necessarily constant. 

\subsubsection{Balancing transformations}
For large $t_\infty$, only the $n$th to $(n+m)$th columns of the transformation matrix $T=U_c\Sigma_cV_c^\dagger V\Sigma^{-1/2}$ must be considered since Eq.~(\ref{eq:repMHsv}) implies that $v_{ci}^\dagger v_{j}=0$ if $n\le i\le n+m$ and $j\notin[n,n+m]$, and hence the rest of $T$ is independent of these $(m+1)$ columns. Using Eq.~(\ref{eq:repMHsv}), these $(m+1)$ columns become:
\begin{eqnarray}
&&\lim_{t_\infty\to+\infty}\left[\begin{array}{ccc}T_{n}&\ldots&T_{n+m}\end{array}\right]\nonumber\\
&&\!\!=\!\!\left[\!\begin{array}{ccc}u_{cn}&\ldots&u_{cn+m}\end{array}\!\!\right]\!\!
\left[\!\!\begin{array}{ccc}\sigma_{cn}&&\\&\!\!\ddots\!\!&\\&&\sigma_{cn+m}\end{array}\!\right]\!\!R_c\!\!
\left[\!\!\begin{array}{ccc}\sigma_{n}^{-1/2}&&\\&\!\!\ddots\!\!&\\&&\sigma_{n+m}^{-1/2}\end{array}\!\right]\!\!.\nonumber
\end{eqnarray}
Therefore, in general, the subspace spanned by the balanced modes $T_i$ for $n\le i \le n+m$ converges to the subspace spanned by ${T_{cn-1}\left[\begin{array}{ccc}p_n&\ldots&p_{n+m}\end{array}\right]}$. Similarly, it can be shown that the adjoint balanced modes $S_i$ for $n\le i \le n+m$ must span the same subspace as ${T_{on-1}\left[\begin{array}{ccc}q_n&\ldots&q_{n+m}\end{array}\right]}$. 
 
\subsubsection{Transformed Gramians}
Finally, since Eq.~(\ref{eq:diagWo}) is still valid here, $W_o$ is diagonalized by $T$ if $V_{(21)}=V_{(11)}$. We again only need to consider the $n$th to $(n+m)$th columns of each matrix since the other columns are unaffected by these modes for large $t_1$ and $t_2$. Using Eq.~(\ref{eq:repMHsv}):
\begin{eqnarray}
&&\left[\begin{array}{ccc}v_{n}&\ldots&v_{n+m}\end{array}\right]_{(21)}=
\left[\begin{array}{ccc}v_{cn}&\ldots&v_{cn+m}\end{array}\right](t_1)R_{c(21)},\nonumber\\
&&\left[\begin{array}{ccc}v_{n}&\ldots&v_{n+m}\end{array}\right]_{(11)}=
\left[\begin{array}{ccc}v_{cn}&\ldots&v_{cn+m}\end{array}\right](t_1)R_{c(11)},\nonumber\\
&&\!\!\Rightarrow\!\!\left[\begin{array}{ccc}v_{n}&\ldots&v_{n+m}\end{array}\right]^\dagger _{(21)}
\left[\begin{array}{ccc}v_{n}&\ldots&v_{n+m}\end{array}\right]_{(11)}\!\!=\!R_{c(21)}^\dagger R_{c(11)}\nonumber.
\end{eqnarray}
In general, $V_{(11)}^\dagger V_{(21)}$ therefore becomes:
\begin{eqnarray}
V_{(11)}^\dagger V_{(21)}&=&\left[\begin{array}{ccc}I&0&0\\
0&R_{c(11)}^\dagger R_{c(21)}&0\\
0&0&I\end{array}\right]\nonumber.
\end{eqnarray}
The transformed Gramian $T(t_1)^\dagger W_o(t_2)T(t_1)$ is thus fully diagonal, except for $(m+1)$ columns which have a dense $((m+1)\times(m+1))$ block along the diagonal. An analogous derivation applied to the controllability Gramian leads to the conclusion that the Gramians are balanced (diagonal and equal), except for the $(m+1)$ columns corresponding to the modes with equal growth rates. The fact that the transformation does not in general balance unstable modes with identical growth rates is not problematic because, as noted above, the subspace spanned by the $(m+1)$ balanced modes does converge, and the rest of the system remains unaffected. In other words, the $(m+1)$ modes are balanced as a whole. As it is undesirable to truncate unstable modes for control purposes, the transformations $T$ and $S$ are still adequate transformations as long as they are computed with a set a snapshots that allows this subspace to converge.

On the other hand, if $R_c$ and $R_o$ {\it were} to tend to constant matrices, the Gramians would become {\it fully balanced} by $T$ and $S$ for large $t_\infty$. We will focus on special cases where this happens in the sections below.

\subsection{Repeated eigenvalues \label{repeigs}}
In this case,  ${\lambda_n=\lambda_{n+1}=\ldots=\lambda_{n+m}=\alpha+i\omega}$. In order to analyse the singular values and vectors of $X$ in this case the transformation is applied to the full controllability Gramian ${W_c(t_\infty)=\int_0^{t_{\infty}}xx^\dagger dt= XX^\dagger }$:
\begin{eqnarray}
T_{cn-1}W_c(t_\infty)T_{cn-1}&=&\int_0^{t_{\infty}}T_{cn-1}xx^\dagger T_{cn-1}dt\nonumber,
\end{eqnarray}
which gives:
\begin{eqnarray}
&\lim_{t_\infty\to+\infty}T_{cn-1}W_c(t_\infty)T_{cn-1}\!=\!\left[\!\!\begin{array}{ccc}u_{cn}&\!\!\ldots\!\!&u_{cn+m}\end{array}\!\!\right]\!\!
\left[\!\!\begin{array}{ccc}\sigma_{cn}^2&&\\&\!\!\ddots\!\!&\\&&\sigma_{cn+m}^2\end{array}\!\right]\!\!
\left[\!\begin{array}{c}u^\dagger _{cn}\\\!\!\vdots\!\!\\u^\dagger _{cn+m}\end{array}\!\right]\!,\nonumber\\
&\!=\!T_{cn-1}\!\left[\!\!\begin{array}{ccc}p_n&\!\!\ldots\!\!&p_{n+m}\end{array}\!\!\right]\!\!
\left[\!\begin{array}{c}q^\dagger _{n}\\\!\!\vdots\!\!\\q^\dagger _{n+m}\end{array}\!\right]\!\!B
B^\dagger \!\!\left[\!\!\begin{array}{ccc}q_n&\!\!\ldots\!\!&q_{n+m}\end{array}\!\!\right]\!\!
\left[\begin{array}{c}p^\dagger _{n}\\\!\!\vdots\!\!\\p^\dagger _{n+m}\end{array}\!\right]\!T_{cn-1}\!
\int_0^{t_{\infty}}\!e^{2\alpha t}dt\label{eq:TcWcTc},
\end{eqnarray}
since ${\lambda+\lambda^*=2\alpha}$. The term to the left of ${\int_0^{t_{\infty}}e^{2\alpha t}dt}$ in Eq.~(\ref{eq:TcWcTc}) is just a constant matrix and the integral itself is a scalar. As a result ${\left[\begin{array}{ccc}u_{cn}&\ldots&u_{cn+m}\end{array}\right]}$ converge for large ${t_\infty}$ and $\int_0^{t_{\infty}}e^{2\alpha t}dt=\left(e^{2\alpha t_{\infty}}-1\right)/\left(2\alpha\right)\approx e^{2\alpha t_{\infty}}/\left(2\alpha\right)$. The singular values can thus be written ${\sigma_{ci}\approx\tilde \sigma_{ci}e^{\alpha t_{\infty}}/\sqrt{2\alpha}}$, where ${\tilde \sigma_{ci}^2}$ are the singular values of the constant part of Eq.~(\ref{eq:TcWcTc}). If ${\alpha =0}$, then the modes are marginally stable and ${\int_0^{t_{\infty}}dt=t_{\infty}}$ so ${\sigma_{ci}\approx\tilde \sigma_{ci}\sqrt{t_{\infty}}}$. Analogous conclusions can be drawn regarding the observability singular values and vectors: ${\left[\begin{array}{ccc}u_{on}&\ldots&u_{on+m}\end{array}\right]}$ converge and ${\sigma_{oi}\approx\tilde \sigma_{oi}e^{\alpha  t_{\infty}}/\sqrt{2\alpha}}$ (and ${\sigma_{oi}\approx\tilde \sigma_{oi}\sqrt{t_{\infty}}}$ for $\alpha =0$). 

The SVD of the matrix $M$ defined in  Eq.~(\ref{eq:Mmatrix}) can thus be written:
\begin{eqnarray}
\lim_{t_\infty\to+\infty}M&=&R_o\left[\begin{array}{ccc}\sigma_{n}&&\\&\ddots&\\&&\sigma_{n+m}\end{array}\right]R_c^\dagger, \nonumber\\
&=&\frac{e^{2\alpha  t_{\infty}}}{2\alpha}R_o\left[\begin{array}{ccc}\tilde \sigma_{n}&&\\&\ddots&\\&&\tilde \sigma_{n+m}\end{array}\right]R_c^\dagger ,\nonumber\\
&=&\frac{e^{2\alpha  t_{\infty}}}{2\alpha}\tilde M,
\end{eqnarray}
where
\begin{eqnarray}
\tilde M&=&
               \left[\begin{array}{c}\tilde \sigma_{on}u_{on}^\dagger \\\vdots\\\tilde \sigma_{on+m}u_{on+m}^{\dagger }\end{array}\right]
               \left[\begin{array}{ccc}u_{cn}\tilde \sigma_{cn}&\ldots&u_{cn+m}\tilde \sigma_{cn+m}\end{array}\right].\nonumber
\end{eqnarray}
Here ${\sigma_i=\tilde \sigma_ie^{2\alpha  t_{\infty}}/(2\alpha)}$ (and ${\sigma_i=\tilde \sigma_it_\infty}$ for $\alpha =0$), and $\tilde M$ tends to a constant matrix (and hence so do its singular values $\tilde \sigma_n$, and vectors $R_o$, $R_c$).

For large $t_\infty$, the $n$th to $(n+m)$th columns of the transformation matrix become:
\begin{eqnarray}
&&\lim_{t_\infty\to+\infty}\left[\begin{array}{ccc}T_{n}&\ldots&T_{n+m}\end{array}\right]\nonumber\\
&&\!=\!\!\left[\!\!\begin{array}{ccc}u_{cn}&\!\!\ldots\!\!&u_{cn+m}\end{array}\!\right]\!\!
 \left[\!\!\begin{array}{ccc}\tilde\sigma_{cn}&&\\&\ddots&\\&&\tilde\sigma_{cn+m}\end{array}\!\right]\!\!R_c\!\!
 \left[\!\!\begin{array}{ccc}\tilde\sigma_{n}^{-1/2}&&\\&\!\!\ddots\!\!&\\&&\tilde\sigma_{n+m}^{-1/2}\end{array}\!\right]\!\!.\nonumber
 \\\nonumber
 \end{eqnarray}
Since $R_c$ and $R_o$ converge to constants for large $t_\infty$, we can conclude that for repeated eigenvalues, the matrices $T$ and $S$ converge and fully balance the Gramians.

\subsection{Complex conjugate eigenvalues}
If again there are $(n-1)$ modes that are more unstable than the complex conjugate pair, then we proceed to identify all of them and 
project them out, as in Sec.~\ref{newbaltrunc}: 
\begin{eqnarray}
&&T_{cn-1}x(t)=T_{cn-1} \left[\!\!\begin{array}{ccc}p_n&\!\!\ldots\!\!&p_{n_x}\end{array}\!\!\right]\!\!
\left[\!\!\begin{array}{ccc}e^{\lambda_nt}&&\\&\!\!\ddots\!\!&\\&&e^{\lambda_{n_x}t}\end{array}\!\!\right]\!\!
\left[\begin{array}{c}q_n^\dagger \\\vdots\\q_{n_x}^{*\dagger }\end{array}\!\!\right]\!\!B,\nonumber\\
&&\Rightarrow\lim_{t\to+\infty}T_{cn-1}x(t)=T_{cn-1} \left[\!\!\begin{array}{cc}p_n&p^*_{n}\end{array}\!\!\right]\!\!
\left[\!\!\begin{array}{cc}e^{\lambda_nt}&0\\0&e^{\lambda_{n}^*t}\end{array}\!\!\right]\!\!
\left[\!\!\begin{array}{c}q_n^\dagger \\q_{n}^{*\dagger }\end{array}\!\!\right]\!\!B.\nonumber
\end{eqnarray}
Finally, this results in:
\begin{eqnarray}
\lim_{t_\infty\to+\infty}T_{cn-1}W_c(t_\infty)T_{cn-1}\!=\!
\!\!\int_0^{t_\infty} T_{cn-1} \left[\!\!\begin{array}{cc}p_n&p^*_{n}\!\!\end{array}\right]\!\!
\left[\begin{array}{cc}q^\dagger BB^\dagger qe^{2\alpha t}     &    q^{\dagger }BB^\dagger q^*e^{2\lambda_n t}     \\    q^{*\dagger }BB^\dagger q e^{2\lambda_n^*t}    &    q^{*\dagger }BB^\dagger q^*e^{2\alpha t} \end{array}\right]\!\!
\left[\begin{array}{c}p_n^\dagger T_{cn-1} \\p_{n}^{*\dagger }T_{cn-1} \end{array}\!\!\right]dt\nonumber.
\end{eqnarray}

\subsubsection{Marginally stable case \label{marginallystable}}
In this case ${e^{2\alpha t}=1}$, and the projected Gramian simplifies to:
\begin{eqnarray}
&&\lim_{t_\infty\to+\infty}T_{cn-1}W_c(t_\infty)T_{cn-1}\nonumber\\
&&=\!t_\infty T_{cn-1}\!\left[\!\!\begin{array}{cc}p_n&p^*_{n}\end{array}\!\!\right]\!\!\!
\left[\!\!\begin{array}{cc}q^\dagger BB^\dagger q     &    q^{\dagger }BB^\dagger q^*\epsilon^+    \\    q^{*\dagger }BB^\dagger q \epsilon^-   &    q^{*\dagger }BB^\dagger q^* \end{array}\!\!\!\right]\!\!\!
\left[\!\!\begin{array}{c}p_n^\dagger T_{cn-1} \\p_{n}^{*\dagger }T_{cn-1} \end{array}\!\!\!\right],\nonumber
\end{eqnarray}
where:
\begin{eqnarray}
\epsilon^{\pm}=\frac{e^{\left(\pm2i\omega_n t_\infty\right)}-1}{\pm2i\omega_nt_\infty}\Rightarrow\lim_{t_\infty\to+\infty}\epsilon^{\pm}=0\nonumber,
\end{eqnarray}
and hence the projected Gramian can be written:
\begin{eqnarray}
&&\lim_{t\to+\infty}T_{cn-1}W_c(t_\infty)T_{cn-1}\nonumber\\
&&=\!t_\infty T_{cn-1} \!\left[\!\!\begin{array}{cc}p_n&p^*_{n}\end{array}\!\!\right]\!\!
\left[\!\!\begin{array}{cc}q^\dagger BB^\dagger q     &    0    \\   0    &    q^{*\dagger }BB^\dagger q^* \end{array}\!\!\right]\!\!
\left[\!\!\begin{array}{c}p_n^\dagger T_{cn-1} \\p_{n}^{*\dagger }T_{cn-1} \end{array}\!\!\right],\nonumber\\
&&=t_\infty \tilde W_c\nonumber,
\end{eqnarray}
where $\tilde W_c$ is constant, and hence has constant singular vectors $\left[\begin{array}{cc}u_{cn}&u_{cn+1}\end{array}\right]$ and singular values that can be written: ${t_\infty \tilde \sigma_{cn}^2}$ and $t_\infty \tilde \sigma_{cn+1}^2$, with ${\tilde \sigma_{cn}}$, ${\tilde \sigma_{cn+1}}$ constant real scalars. The observability Gramian behaves analogously. As a consequence, the matrix $M$ defined in Eq.~(\ref{eq:Mmatrix}) can be written $M=t_\infty\tilde M$, where $\tilde M$ is constant, so $R_c$ and $R_o$ tend to constants and $\sigma_i=t_\infty\tilde \sigma_i$, where $\tilde \sigma_i$ are the constant singular values of $\tilde M$, just like in Sec.~\ref{repeigs}. The conclusions that $T$ and $S$ tend to constant matrices that fully balance the Gramians follow in the same way.

\subsubsection{Unstable case \label{unstableccpair}}
If $\alpha>0$, we obtain:
\begin{eqnarray}
&&\lim_{t\to+\infty}T_{cn-1}W_c(t_\infty)T_{cn-1}\nonumber\\
&&=\!\!\frac{ T_{cn-1}\!\! \left[\!\!\begin{array}{cc}p_n&p^*_{n}\end{array}\!\!\right]\!\!
\left[\!\!\begin{array}{cc}q^\dagger BB^\dagger q     &    q^{\dagger }BB^\dagger q^*\epsilon^+    \\    q^{*\dagger }BB^\dagger q \epsilon^-   &    q^{*\dagger }BB^\dagger q^* \end{array}\!\!\right]\!\!
\left[\!\!\begin{array}{c}p_n^\dagger T_{cn-1} \\p_{n}^{*\dagger }T_{cn-1} \end{array}\!\!\right]}{2\alpha\,/\,\left(e^{2\alpha t_\infty}-1\right) }. \nonumber
\end{eqnarray}
Unlike the marginally stable case however:
\begin{eqnarray}
\epsilon^{\pm}&=&\left(\frac{\alpha}{\alpha\pm i\omega_n}\right)\frac{e^{2\left(\alpha \pm i\omega_n\right)t_\infty}-1}{e^{2\alpha t_\infty}-1},\nonumber\\
\Rightarrow\lim_{t_\infty\to+\infty}\epsilon^{\pm}&=&\left(\frac{\alpha}{\alpha\pm i\omega_n}\right)e^{2i\omega_n t_\infty}\neq0\nonumber.
\end{eqnarray}
Therefore, even for large $t_\infty$ the left controllability (observability) singular vectors oscillate but stay in the two-dimensional plane defined by the eigenvectors. Similarly the Hankel singular values grow on average exponentially as $e^{2\alpha t_\infty}$, but the growth rate also oscillates around this mean trend. As discussed in Sec.~\ref{appxgeneral}, for each unstable complex conjugate pair of modes, two columns of $T$ and two rows of $S$ will not converge regardless of $t_\infty$ (but will stay bounded), and a full $(2\times2)$ block will be present along the diagonal of the transformed Gramians, while the rest of the matrix will be independently balanced. If $t_\infty$ is large enough for the subspace spanned by the two corresponding columns of $T$ (and rows of $S$) to be converged, then any $T$ and $S$ transformations can therefore be used an adequate balancing transformations, where the complex conjugate pair is considered to be balanced as a whole.


\section{Induction proof for the Hankel matrix SVD \label{hankinduction}}
As in {\it Proposition 1} in Sec.~\ref{newbaltrunc}, we will use a proof by induction to show that for large $t_\infty$, the singular vectors and values corresponding to unstable modes of the Hankel Matrix ${M_H=Z^\dagger X=U\Sigma V^\dagger =V_o\Sigma_oU_o^\dagger U_c\Sigma_cV_c^\dagger }$ tend to:
\begin{eqnarray}
\begin{cases}
u_i=v_{oi},\\
\sigma_i=\sigma_{oi}u_{oi}^\dagger u_{ci}\sigma_{ci},\\
v_i= v_{ci} \nonumber.
\end{cases}
\end{eqnarray}
We choose the induction hypothesis $({\cal I}_n)$ to be the fact that the proof holds for all $i\le n$. For the base case $({\cal I}_1)$, if the first mode is unstable:
\begin{eqnarray}
\lim_{t_\infty\to+\infty} M_H&=&v_{o1}\sigma_{o1}u_{o1}^\dagger u_{c1}\sigma_{c1}v_{c1}^\dagger =u_1\sigma_1v_1^\dagger \nonumber,
\end{eqnarray}
since, if $t_\infty$ is large, ${\sigma_{c1}\gg\sigma_{ci}}$ and ${\sigma_{o1}\gg\sigma_{oi}}$ for $i>1$. $({\cal I}_{1})$ therefore holds: $u_1=v_{o1}$, $\sigma_1=\sigma_{o1}u_{o1}^\dagger u_{c1}\sigma_{c1}$ and $v_1= v_{c1}$.

In the inductive step, we prove that if $({\cal I}_n)$ holds for some rank $n$, then $({\cal I}_{n+1})$ also holds. The transformation matrices $T_{li}$ defined in Eq.~(\ref{eq:TliTri}) project out all the left singular vectors $u_{j}$ for all $j\le i$, i.e. $T_{li}u_{j}=0$, $T_{lj}u_{i}=u_{i}$. $T_{ri}$ acts in an analogous way on the right singular vectors. As we are assuming $({\cal I}_n)$ to hold for rank $n$:
\begin{eqnarray}
&&T_{ln}M_HT_{rn}\!=\!\left[\!\!\begin{array}{ccc}u_{n+1}&\ldots&u_{n_x}\end{array}\right]\!\!
\left[\!\!\begin{array}{ccc}\sigma_{n+1}&&\\&\!\!\ddots\!\!&\\&&\sigma_{n_x}\end{array}\!\right]\!\!
\left[\!\!\begin{array}{c}v^\dagger _{n+1}\\\vdots\\v^\dagger _{n_x}\end{array}\!\right],\nonumber\\
&&=T_{ln}\left[\begin{array}{c}v_{on+1}^\dagger \\\vdots\\v_{on_x}^\dagger \end{array}\!\right]^\dagger \!
\left[\!\!\begin{array}{c}\sigma_{on}u^\dagger _{on+1}\\\vdots\\\sigma_{on_x}u^\dagger _{on_x}\end{array}\!\right]\!\!
\left[\!\!\begin{array}{c}\sigma_{cn}u_{cn+1}^\dagger \\\vdots\\\sigma_{cn_x}u_{cn_x}^\dagger \end{array}\!\right]^\dagger \!
\left[\!\!\begin{array}{c}v^\dagger _{cn+1}\\\vdots\\v^\dagger _{cn_x}\end{array}\!\right]\!\!T_{rn}\nonumber.
\end{eqnarray}
As a result, if the $(n+1)$th mode is unstable:
\begin{eqnarray}
&&\lim_{t_\infty\to+\infty}T_{ln}M_HT_{rn}\nonumber\\
&&=\left(T_{ln}v_{on+1}\right)\left(\sigma_{on+1}u_{on+1}^\dagger u_{cn+1}\sigma_{cn+1}\right)\left(T_{rn}v_{cn+1}\right)^\dagger \nonumber,
\end{eqnarray}
because if $t_\infty$ is large, ${\sigma_{ci}\gg\sigma_{cj}}$ and ${\sigma_{oi}\gg\sigma_{oj}}$ for $i<j$. Since $V_o$ and $V_c$ are individually orthonormal bases, $v_{on+1}$ is normal to $v_{oi}=u_{i}$ if $i\le n$ and $v_{cn+1}$ is normal to $v_{ci}=v_{i}$ if $i\le n$:
\begin{eqnarray}
&&\lim_{t_\infty\to+\infty}\!\! T_{ln}M_HT_{rn}
\!= \!v_{on+1}\!\!\left(\!\sigma_{on+1}u_{on+1}^\dagger u_{cn+1}\sigma_{cn+1}\!\right)\!v_{cn+1}^\dagger\nonumber,
\end{eqnarray}
and therefore $({\cal I}_{n+1})$ holds too: $u_{n+1}=v_{on+1}$, ${\sigma_{n+1}=\sigma_{on+1}u_{on+1}^\dagger u_{cn+1}\sigma_{cn+1}}$ and $v_{n+1}= v_{cn+1}$. This completes the inductive step and, along with the base step, this concludes the proof by induction.


\bigskip
\section{Proof of the convergence of the ratio of singular values  \label{cstsigratios}}
In this section, we prove that $\sigma_{ci}^2/\sigma_i$ tends to a constant for large $t_\infty$.
\begin{eqnarray}
\lim_{t_\infty\to+\infty}\frac{\sigma^2_{ci}}{\sigma_{i}}&=&\frac{\sigma^2_{ci}}{\sigma_{oi}u_{oi}^\dagger u_{ci}\sigma_{ci}}
=\frac{\sigma_{ci}}{\sigma_{oi}u_{oi}^\dagger u_{ci}},\nonumber\\
&=&\frac{\lVert T_{ci-1}p_i \rVert\lVert \beta_{i} \rVert}{\lVert T_{oi-1}q_i \rVert\lVert \xi_{i} \rVert}\frac{1}{u_{oi}^\dagger u_{ci}}\nonumber.
\end{eqnarray}
Only ${\lVert \beta_{i} \rVert}$ and ${\lVert \xi_{i} \rVert}$ are not constants for large $t_\infty$ in the above expression and:
\begin{eqnarray}
 \lVert \beta_{i} \rVert^2&=&\beta_i^\dagger \beta_i,\nonumber\\
 &=&\left[\!\!\begin{array}{c}\sqrt{\delta_{c1}}B^\dagger q_ie^{\left(\lambda^*_it_{c1}\right)}\\\vdots\\\sqrt{\delta_{cN_c}}B^\dagger q_ie^{\left(\lambda^*_it_{cN_c}\right)}\end{array}\!\!\right]^\dagger 
 \!\!
 \left[\!\!\begin{array}{c}\sqrt{\delta_{c1}}B^\dagger q_ie^{\left(\lambda^*_it_{c1}\right)}\\\vdots\\\sqrt{\delta_{cN_c}}B^\dagger q_ie^{\left(\lambda^*_it_{cN_c}\right)}\end{array}\!\!\right]\nonumber,\\
 &=&q_i^\dagger BB^\dagger q_i\sum_{k=1}^{N_c}e^{\left(2\alpha_it_{ck}\right)}\delta_{ck}.
 \end{eqnarray}
 We can therefore use the approximation:
 \begin{eqnarray}
 \lVert \beta_{i} \rVert^2&\approx& q_i^\dagger BB^\dagger q_i\int_{0}^{t_{\infty}}e^{\left(2\alpha_it\right)}dt,\nonumber\\
 &=&\lVert B^\dagger q_i\rVert^2\int_{0}^{t_{\infty}}e^{\left(2\alpha_it\right)}dt,\nonumber
\end{eqnarray}
where we have defined ${\lambda_i+\lambda_i^*=2\alpha_i}$ and ${\alpha_i\in\mathbb{R}}$. Similarly:
\begin{eqnarray}
 \lVert \xi_{i} \rVert^2&=&p_i^\dagger C^\dagger Cp_i\sum_{k=1}^{N_o}e^{\left(2\alpha_it_{ok}\right)}\delta_{ok},\nonumber\\
 &\approx& \lVert Cp_i\rVert^2\int_{0}^{t_{\infty}}e^{\left(2\alpha_it\right)}dt.\nonumber
\end{eqnarray}
Therefore:
\begin{eqnarray}
\lim_{t_\infty\to+\infty}\frac{\sigma^2_{ci}}{\sigma_{i}}&=&\frac{\lVert T_{ci-1}p_i \rVert}{\lVert T_{oi-1}q_i \rVert}\frac{1}{u_{oi}^\dagger u_{ci}}\frac{\lVert B^\dagger q_i\rVert}{\lVert Cp_i\rVert}\nonumber,
\end{eqnarray}
which is a constant.

\bibliographystyle{acm}
\bibliography{FlinoisMorgansSchmid.bib}

\end{document}